\documentclass[10pt]{article}
\usepackage{epsf}
\usepackage{amsmath}

\allowdisplaybreaks

\usepackage[showframe=false]{geometry}
\usepackage{changepage}

\usepackage{epsfig}
\usepackage{amssymb}

\usepackage{amsthm}
\usepackage{setspace}
\usepackage{cite}
\usepackage{mcite}

\usepackage{algorithmic}  
\usepackage{algorithm}

\usepackage{shadow}
\usepackage{fancybox}
\usepackage{fancyhdr}

\usepackage{color}
\usepackage[usenames,dvipsnames,svgnames,table]{xcolor}

\definecolor{mypurple}{rgb}{.4,.0,.5}

\usepackage{xcolor}

\usepackage{color}

\definecolor{darkgreen}{rgb}{0, 0.4,0}

\definecolor{purplebrown}{rgb}{0.5,0.1,0.6}

\definecolor{ultclupcol}{rgb}{0.1,0.5,0.5}
\newcommand{\ultclupcol}[1]{\textcolor{ultclupcol}{#1}}

\newcommand{\bl}[1]{\textcolor{blue}{#1}}

\newcommand{\prp}[1]{\textcolor{purple}{#1}}

\definecolor{shadebrown}{rgb}{0.1,0.1,0.9}
\definecolor{lightblue}{rgb}{0.2,0,1}

\usepackage{fancybox}
\usepackage{graphicx}
\usepackage{epstopdf}
\usepackage{epsfig}
\usepackage{wrapfig}
\usepackage{subfigure}

\usepackage{xcolor}

\usepackage{tcolorbox}
\tcbuselibrary{skins}

\newtcbox{\xmybox}{on line,
arc=7pt,
before upper={\rule[-3pt]{0pt}{10pt}},boxrule=0pt,
boxsep=0pt,left=6pt,right=6pt,top=0pt,bottom=0pt,enhanced, coltext=blue, colback=white!10!yellow}

\newtcbox{\xmyboxa}{on line,
arc=7pt,
before upper={\rule[-3pt]{0pt}{10pt}},boxrule=0pt,
boxsep=0pt,left=6pt,right=6pt,top=0pt,bottom=0pt,enhanced, colback=white!10!yellow}

\newtcbox{\xmyboxb}{on line,
arc=7pt,
before upper={\rule[-3pt]{0pt}{10pt}},boxrule=1pt,colframe=darkgreen!100!blue,
boxsep=0pt,left=6pt,right=6pt,top=0pt,bottom=0pt,enhanced, colback=white!10!yellow}

\newtcbox{\xmyboxc}{on line,
arc=7pt,
before upper={\rule[-3pt]{0pt}{10pt}},boxrule=.7pt,colframe=blue!100!blue,
boxsep=0pt,left=6pt,right=6pt,top=0pt,bottom=0pt,enhanced, coltext=blue, colback=white!10!yellow}

\newtcbox{\xmytboxa}{on line,
arc=7pt,
before upper={\rule[-3pt]{0pt}{10pt}},boxrule=.0pt,colframe=pink!50!yellow,
boxsep=0pt,left=6pt,right=6pt,top=0pt,bottom=0pt,enhanced, coltext=white, colback=blue!40!red}

\newtcbox{\xmytboxb}{on line,
arc=7pt,
before upper={\rule[-3pt]{0pt}{10pt}},boxrule=.0pt,colframe=pink!50!yellow,
boxsep=0pt,left=6pt,right=6pt,top=0pt,bottom=0pt,enhanced, coltext=white, colback=white!40!green}

\usepackage[hyphens]{url}

\usepackage[colorlinks=true,
            linkcolor=black,
            urlcolor=blue,
            citecolor=purple]{hyperref}

\usepackage{breakurl}

\def\y{{\bf y}}
\def\v{{\bf v}}
\def\x{{\bf x}}

\def\x{{\mathbf x}}

\def\v{{\bf v}}
\def\x{{\bf x}}
\def\y{{\bf y}}

\def\be{\begin{equation}}
\def\ee{\end{equation}}
\def\ba{\left[\begin{array}}
\def\ea{\end{array}\right]}

\def\v{{\bf v}}
\def\x{{\bf x}}
\def\y{{\bf y}}

\def\1{{\bf 1}}

\def\0{{\bf 0}}

\def\erfinv{\mbox{erfinv}}

\def\mR{{\mathbb R}}

\def\mE{{\mathbb E}}

\newtheorem{theorem}{Theorem}
\newtheorem{corollary}{Corollary}

\begin{document}

\begin{singlespace}

\title {Rephased CLuP 
}
\author{
\textsc{Mihailo Stojnic
\footnote{e-mail: {\tt flatoyer@gmail.com}} }}
\date{}
\maketitle

\centerline{{\bf Abstract}} \vspace*{0.1in}

In \cite{Stojnicclupint19,Stojnicclupcmpl19,Stojnicclupplt19} we introduced CLuP, a \bl{\textbf{Random Duality Theory (RDT)}} based algorithmic mechanism that can be used for solving hard optimization problems. Due to their introductory nature, \cite{Stojnicclupint19,Stojnicclupcmpl19,Stojnicclupplt19} discuss the most fundamental CLuP concepts. On the other hand, in our companion paper \cite{Stojniccluplargesc20} we started the story of going into a bit deeper details that relate to many of other remarkable CLuP properties with some of them reaching well beyond the basic fundamentals. Namely, \cite{Stojniccluplargesc20} discusses how a somewhat silent RDT feature (its algorithmic power) can be utilized to ensure that CLuP can be run on very large problem instances as well. In particular, applying CLuP to the famous MIMO ML detection problem we showed in \cite{Stojniccluplargesc20} that its a large scale variant, $\text{CLuP}^{r_0}$, can handle with ease problems with \textbf{\emph{several thousands}} of unknowns with theoretically minimal complexity per iteration (only a single matrix-vector multiplication suffices). In this paper we revisit MIMO ML detection and discuss another remarkable phenomenon that emerges within the CLuP structure, namely the so-called \bl{\textbf{\emph{rephasing}}}. As MIMO ML enters the so-called low $\alpha$ regime (fat system matrix with ratio of the number of rows and columns, $\alpha$, going well below $1$) it becomes increasingly difficult even for the basic standard CLuP to handle it. However, the discovery of the rephasing ensures that CLuP remains on track and preserves its ability to achieve the ML performance. To demonstrate the power of the rephasing we also conducted quite a few numerical experiments, compared the results we obtained through them to the theoretical predictions, and observed an excellent agreement.

\vspace*{0.25in} \noindent {\bf Index Terms: Rephasing; Large scale CLuP; ML - detection; MIMO systems; Algorithms; Random duality theory}.

\end{singlespace}

\section{Introduction}
\label{sec:back}

We introduced the fundamentals of the so-called \bl{\textbf{CLuP}} optimization concept in a series of recent papers \cite{Stojnicclupint19,Stojnicclupcmpl19,Stojnicclupplt19}. As was hinted on many occasions in \cite{Stojnicclupint19,Stojnicclupcmpl19,Stojnicclupplt19} CLuP is a very powerful tool that can be used for solving hard optimization problems. To demonstrate its efficiency we chose for the introductory considerations the famous, so-called MIMO ML detection problem. A quite a few CLuP features immediately distinguished themselves. We will here single out the two of them: 1) first, it was clear that CluP can achieve the so-called ML level of performance basically with an ease and for fairly large problems; 2) second, it could attack computationally the most challenging regimes, in particular the so-called low $\alpha$ regimes where $\alpha$ -- the ratio of the number of rows and the number of columns of the system matrix -- goes well below $1$. In our companion paper \cite{Stojniccluplargesc20} we went a bit further regarding the first of these features and provided a thorough discussion regarding the large scale CLuP capabilities. Along the same lines we designed a particularly tailored large-scale version of the CLuP, called $\text{CLuP}^{r_0}$, and showed that it can easily handle problems of sizes of few thousands. Moreover, given that it has a theoretically minimal quadratic complexity per iteration (which includes basically only a single matrix-vector multiplication), it is expected to be an excellent tool in the big data era where ability to work with the dimensions of few tens/hundreds of thousands or millions is particularly desirable. In this paper we will discuss in bit more details the second of the above CLuP features, namely its ability to handle the so-called low $\alpha$ regimes. Before we reach the point to understand how important is to be able to handle such regimes and how difficult task such a handling can be, we will need a bit of problem introduction.

As in \cite{Stojnicclupint19,Stojnicclupcmpl19,Stojnicclupplt19} and more recently \cite{Stojniccluplargesc20}, we will here also choose MIMO ML detection problem as the benchmark for showcasing CLuP's capabilities. We mention though right here at the beginning that all the concepts that we will present below (as well as those that we presented in \cite{Stojniccluplargesc20} and earlier in \cite{Stojnicclupint19,Stojnicclupcmpl19,Stojnicclupplt19}) are very generic and in no way restricted to only MIMO ML. However, to achieve a large degree of parallelism with the \cite{Stojnicclupint19,Stojnicclupcmpl19,Stojnicclupplt19,Stojniccluplargesc20} and to facilitate both the presentation of the main ideas and their following we chose the same benchmark MIMO ML detection problem as we did in \cite{Stojnicclupint19,Stojnicclupcmpl19,Stojnicclupplt19,Stojniccluplargesc20}.

Since we already introduced the MIMO ML detection problem on multiple occasions in \cite{Stojnicclupint19,Stojnicclupcmpl19,Stojnicclupplt19,Stojniccluplargesc20} we will below provide only the most fundamental definitions needed for further following and refer the interested reader to our earlier papers for a more complete, details-filled picture. Also, while paralleling the presentations from \cite{Stojnicclupint19,Stojnicclupcmpl19,Stojnicclupplt19,Stojniccluplargesc20} will be in the interest of easing the following, avoiding repeating a tone of details already provided in these papers and focusing instead on the most important differences will be in the interest of reemphasizing the key points that we want to present here. As usual, we start with a brief description of the linear MIMO system corrupted with noise. Such systems are most often described through the following analytical form
\begin{eqnarray}\label{eq:linsys1}
\y=A\x_{sol}+\sigma\v.
\end{eqnarray}
As in \cite{Stojnicclupint19,Stojnicclupcmpl19,Stojnicclupplt19,Stojniccluplargesc20}, $\y\in\mR^m$ is the output of the system, $\x_{sol}\in\mR^n$ is the input of the system, $A\in\mR^{m\times n}$ is the system matrix that models how the input and the output of the system are connected, $\v\in\mR^m$ is the noise vector, and $\sigma$ is a noise scaling factor that will be helpful in defining and controlling the signal-to-noise (SNR) ratio. Of course, it is rather well known that quite a few areas including, for example, signal processing, machine learning, statistics, and linear estimation utilize these types of models as some of their most important technical tools.

Following further \cite{Stojnicclupint19,Stojnicclupcmpl19,Stojnicclupplt19,Stojniccluplargesc20}, we will here also be interested in the recovery of the $\x_{sol}$. Moreover, similarly to \cite{Stojnicclupint19,Stojnicclupcmpl19,Stojnicclupplt19,Stojniccluplargesc20}, we will consider the so-called coherent detection, i.e. the detection where the matrix $A$ is assumed as known at the system's output. Also, as in  \cite{Stojnicclupint19,Stojnicclupcmpl19,Stojnicclupplt19,Stojniccluplargesc20}, we will here be interested in the so-called linear-statistical regimes. Namely, we will assume that system dimensions are such that $m=\alpha n$ with $\alpha>0$ and $n$ and $m$ large. A bit differently from \cite{Stojnicclupint19,Stojnicclupcmpl19,Stojnicclupplt19,Stojniccluplargesc20}, here we will be mostly interested in scenarios where $\alpha$ is well below $1$. As for the statistical aspects, we will again consider the standard Gaussian setup with the elements of $A$ and $\v$ being i.i.d. standard normals (a bit more on the generality of such setup can be found in \cite{Stojnicclupint19,Stojnicclupcmpl19,Stojnicclupplt19,Stojniccluplargesc20} and on various occasions earlier in \cite{StojnicCSetam09,StojnicCSetamBlock09,StojnicISIT2010binary,StojnicDiscPercp13,StojnicUpper10,StojnicGenLasso10,StojnicGenSocp10,StojnicPrDepSocp10,StojnicRegRndDlt10,Stojnicbinary16fin,Stojnicbinary16asym}). As in \cite{Stojnicclupint19,Stojnicclupcmpl19,Stojnicclupplt19,Stojniccluplargesc20}, $\x_{sol}$ will be binary (such an assumption is of course not necessary as all concepts that we will present below will be in place for almost any $\x_{sol}$; quite a few different other options for $\x_{sol}$ we will discuss in separate papers).

Finally we are in position to define the key optimization problem behind the entire ML detection concept
\begin{eqnarray}\label{eq:ml1}
\hat{\x}=\min_{\x\in{\cal X}}\|\y-A\x\|_2,
\end{eqnarray}
where ${\cal X}$ is the set of all allowed vectors $\x_{sol}$ and since we will impose no other restriction on the above mentioned binary assumption we will also have ${\cal X}=\{-\frac{1}{\sqrt{n}},\frac{1}{\sqrt{n}}\}^n$.

It is of course well known that the problem in (\ref{eq:ml1}) when viewed within the frame of the classical complexity theory is NP hard. Still quite a few techniques have been developed over last several decades that are more or less successful in solving it. We refer to \cite{Stojnicclupint19,Stojnicclupcmpl19,Stojnicclupplt19,Stojniccluplargesc20} for a bit of more extensive discussion regarding the most relevant prior work. Here we just briefly mention that the most well known heuristics are based on the so-called convex relaxation techniques with the Ball, Polytope, and SDP convex relaxations being the most prominent (more on some of these can be found in e.g. \cite{GolVanLoan96Book,GroLovSch93Book,vanMaarWar00,GoeWill95}). Contrary to heuristics, one is often interested in exact solutions of (\ref{eq:ml1}). When it comes to solving (\ref{eq:ml1}) exactly then the so-called Sphere-decoder (SD) algorithm from \cite{FinPhoSD85,HassVik05,JalOtt05} and the Branch-and-bound algorithms from \cite{StojnicBBSD08,StojnicBBSD05} are from the mathematical point of view among the most desirable.

The above mentioned \bl{\textbf{CLuP}} that we introduced in \cite{Stojnicclupint19,Stojnicclupcmpl19,Stojnicclupplt19} through a systematic \bl{\textbf{Random Duality Theory (RDT)}} type of analysis relying on a long line of our earlier work \cite{StojnicCSetam09,StojnicCSetamBlock09,StojnicISIT2010binary,StojnicDiscPercp13,StojnicUpper10,StojnicGenLasso10,StojnicGenSocp10,StojnicPrDepSocp10,StojnicRegRndDlt10,Stojnicbinary16fin,Stojnicbinary16asym}, is a new alterative that provides quite a few desirable features. Of course, one of the most dominant is a very favorable combination of its speed and exactness. Namely, it achieves the speed of the convex relaxation heuristics while maintaining the recovery precision of the algorithms from \cite{FinPhoSD85,HassVik05,JalOtt05,StojnicBBSD08,StojnicBBSD05}. Moreover, as demonstrated in \cite{Stojniccluplargesc20}, CLuP can be implemented as a large scale mechanism that can handle sizes of problems virtually untouchable by any of the techniques from \cite{FinPhoSD85,HassVik05,JalOtt05,StojnicBBSD08,StojnicBBSD05}. Another of the CLuP's features that was particularly emphasized in \cite{Stojnicclupint19,Stojnicclupcmpl19,Stojnicclupplt19} is its particular efficiency in handling problems where $\alpha$ is below $1$. Those scenarios are even more difficult for algorithms from \cite{FinPhoSD85,HassVik05,JalOtt05,StojnicBBSD08,StojnicBBSD05}. While we did put an emphasis on $\alpha<1$  regime in \cite{Stojnicclupint19,Stojnicclupcmpl19,Stojnicclupplt19}, we basically mostly worked in a so-called moderately low $\alpha$ regime. Namely, we chose $\alpha=0.8$ and CLuP was able to handle it pretty much on the ML level in all of the important regimes. Below, we will see that as $\alpha$ continues to go down even the basic CLuP can have a bit of trouble getting to the ML level of performance. However, we will uncover another remarkable phenomenon inherently nested within the CLuP structure that will enable circumventing such troubles and ensuring that CLuP can maintain its superiority in achieving the ML level of performance in these regimes as well. There are however quite a few technicalities that we will need to address/introduce before we can discuss these particularities in greater details.

The presentation will consists of several main parts. We start below by recalling on the basic and large scale CLuP fundamentals. We will then discuss the relation between the ML and CLuP and how it can effect the design of efficient algorithms. Once we are equipped with all these technicalities we will discuss some of the key concepts that enable CLuP to remain efficient even in computationally hardest regimes. All our considerations will be in parallel accompanied with quite a few results obtained through numerical experiments as well. Finally, at the end we will provide a brief summary of everything that we present below. As we stated above (and as we will state on quite a few occasions below), we will often assume a decent level of familiarity with our earlier works (most particular a familiarity with \cite{Stojnicclupint19,Stojnicclupcmpl19,Stojnicclupplt19,Stojniccluplargesc20} and to a degree a familiarity with \cite{StojnicCSetam09,StojnicCSetamBlock09,StojnicISIT2010binary,StojnicDiscPercp13,StojnicUpper10,StojnicGenLasso10,StojnicGenSocp10,StojnicPrDepSocp10,StojnicRegRndDlt10,Stojnicbinary16fin,Stojnicbinary16asym}). Along the same lines, we will as often as possible try to avoid repeating many of the details that are already presented in these papers. Sometimes when the context requires we will also reemphasize some of the key concepts that these papers introduced. In general though, the emphasis will be on the main differences and particularities that relate to the problems of interest here.

\section{\bl{CLuP} fundamentals}
\label{sec:clupfund}

In this section we recall on some of the most important theoretical and practical principles on which CLuP relies. We start with recalling on the structure of both, basic and large-scale CLuP and continue with quite a few other intricacies that will be needed for what is our primary goal in this paper.

\subsection{Basic and large scale \bl{CLuP}}
\label{sec:clup}

In \cite{Stojnicclupint19,Stojnicclupcmpl19,Stojnicclupplt19} we introduced the basic CLuP mechanism. It assumes a very simple iterative procedure that can be summarized in the following way: let $\x^{(0)}$ (generated either randomly or chosen deterministically from set ${\cal X}=\{-\frac{1}{\sqrt{n}},\frac{1}{\sqrt{n}}\}^n$) be the starting estimate for $\x_{sol}$ and let $\x^{(i)},i>0$ be defined as
\begin{eqnarray}
\x^{(i+1)}=\frac{\x^{(i+1,s)}}{\|\x^{(i+1,s)}\|_2} \quad \mbox{with}\quad \x^{(i+1,s)}=\mbox{arg}\min_{\x} & & -(\x^{(i)})^T\x  \nonumber \\
\mbox{subject to} & & \|\y-A\x\|_2\leq r\nonumber \\
&& \x\in \left [-\frac{1}{\sqrt{n}},\frac{1}{\sqrt{n}}\right ]^n. \label{eq:clup1}
\end{eqnarray}
The choice of the so-called radius $r$ is one of the key principles that enables CLuP's success. To ensure the easiness of following the main ideas we in \cite{Stojnicclupint19,Stojnicclupcmpl19,Stojnicclupplt19} $r$ was introduced as a multiple of $r_{plt}$, i.e. $r=r_{sc}r_{plt}\sqrt{n}$ where $r_{plt}$ would be the radius that corresponds to the above mentioned polytope relaxation of (\ref{eq:ml1}). That effectively  also ensured thatchossing $r$ is pretty much fully controlled by the choice of its a scaling version $r_{sc}$. A detailed analysis regarding the choice of $r$ or $r_{sc}$ is provided in \cite{Stojnicclupint19,Stojnicclupcmpl19,Stojnicclupplt19}. Moreover, we accompanied all aspects of the theoretical analysis we a large set of numerical experiments that confirmed all theoretical predictions even on moderate systems dimensions of few hundreds. As we have already mentioned, among many interesting favorable CLuP properties, two are particularly useful and relevant with respect to what we will present in this paper. Namely, CLuP's overall computational complexity is on the level of convex relaxation heuristics while its exactness easily approaches the ideal ML level. Moreover, somewhat paradoxically, it was fairly clear as well that CLuP tends to perform even better as the dimensions grow, both, in terms of accuracy as well as in terms of overall scaling complexity being almost exactly equal to the corresponding one of the convex relaxation techniques. Also, both of these features were present in $\alpha<1$ regimes where all available techniques are known to struggle.

In our companion paper \cite{Stojniccluplargesc20}, we went a step further and highlighted the fact that CLuP might have even greater potential as $n$ grows. Namely, we showed how the algorithmic aspects of Random Duality Theory can be utilized to develop a large-scale CLuP. We referred to such a CLuP, as $\text{CLuP}^{r_0}$ and established it as the following contraction principle
\begin{equation}
\x^{(i+1,r)}=\frac{c_{q,2}\x^{(i)}+\hat{\gamma}_1A^T\y\sqrt{\hat{c}_2}-\hat{\gamma}_1A^TA\x^{(i)}\sqrt{\hat{c}_2}}{c_{q,2}-r_{sc}r_{plt}\sqrt{n}},
\x^{(i+1)}=\begin{cases}
  -\frac{1}{\sqrt{n}}, & \mbox{if } \x^{(i+1,r)}\leq -\frac{1}{\sqrt{n}} \\
  \x^{(i+1,r)}, & \mbox{if } -\frac{1}{\sqrt{n}}\leq \x^{(i+1,r)}\leq \frac{1}{\sqrt{n}} \\
  \frac{1}{\sqrt{n}}, & \mbox{otherwise},
\end{cases}\label{eq:LSclup6}
\end{equation}
where determining all critical parameters is discussed throughout the analysis of the machinery presented in \cite{Stojniccluplargesc20}. As the theoretical analysis predicted and numerous numerical simulations confirmed, the above $\text{CLuP}^{r_0}$ turned out to be indeed a very powerful mechanism. We were able to easily solve problem instances with $n$ being of the order of several thousands. Moreover, it was immediately clear that not only can we solve such problem instances but one can not design an algorithm to attack this binary MIMO ML that would have a lower per iteration scaling complexity. That of course immediately implied that the above procedure is particularly well tailored to handle dimensions of tens/hundreds of thousands or millions which are expected to dominate in the future big data era. Of course, all of this turned out to be possible while maintaining excellent ability to handle problems of moderately small dimensions of a few hundreds as well. Such a dual ability to be able to cover both large scale as well as moderately small problems is typically very rarely seen and usually fairly hard to achieve in any at least somewhat challenging optimization considerations, let alone in those like the MIMO ML where the classical complexity theory predicts non-polynomial hardness barrier.

\begin{figure}[htb]
\centering
\centerline{\epsfig{figure=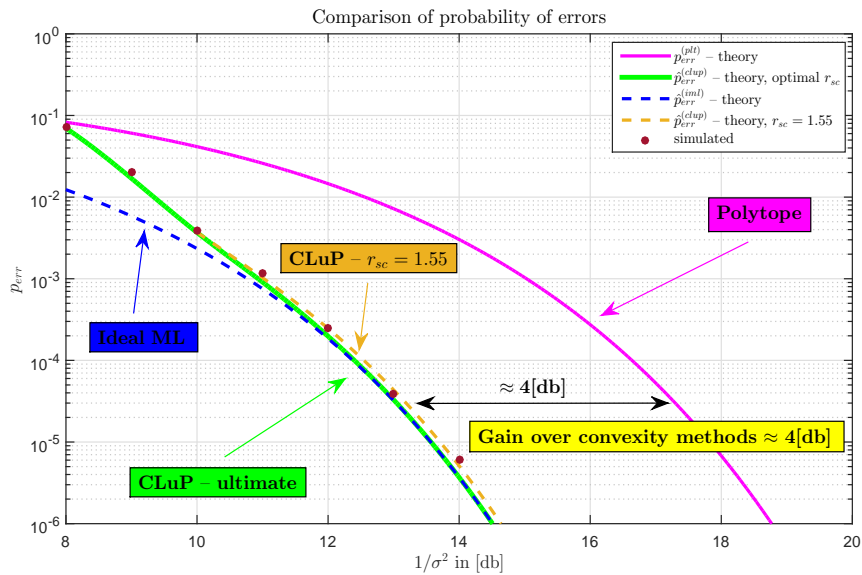,width=13.5cm,height=8cm}}
\caption{Comparison of $p_{err}$ as a function of $1/\sigma^2$; $\alpha=0.8$; $n=2000$}
\label{fig:HLsimLSclupprerr}
\end{figure}

In Figure \ref{fig:HLsimLSclupprerr} we recall on the performance of $\text{CLuP}^{r_0}$ (we refer the interested reader to \cite{Stojniccluplargesc20} for the details and a bit more complete explanations related to the practical technicalities of the implementations presented in the figure).
Here, we just briefly observe that the problem dimension is already not that small any more ($n=2000$) and the accuracy is such that the standard polytope relaxation remains about $4$[db] behind while the so-called ideal ML performance is pretty much achieved. Of course, a particular emphasis is on the fact that $\alpha=0.8$, in other words all these features CLuP possesses even for $\alpha<1$.

All of the above seems very promising and one naturally wonders if the CLuP maintains such abilities as $\alpha$ goes down even further. A discussion in this direction is the main topic of the remainder of this paper. Below, we will first highlight what kind of difficulties one faces as $\alpha$ goes down and how CLuP can handle the challenges that such difficulties are posing. As one of our ultimate goals is achieving the ML performance, we will follow the path that we designed in our introductory CLuP paper \cite{Stojnicclupint19}. Namely, we will first examine the ML performance itself and what kind of technical intricacies such a performance has within itself and then try to reconnect them to the CLuP.

\subsection{Revisiting the ML in low $\alpha$ regime}
\label{sec:mllowalpha}

\begin{figure}[htb]
\centering
\centerline{\epsfig{figure=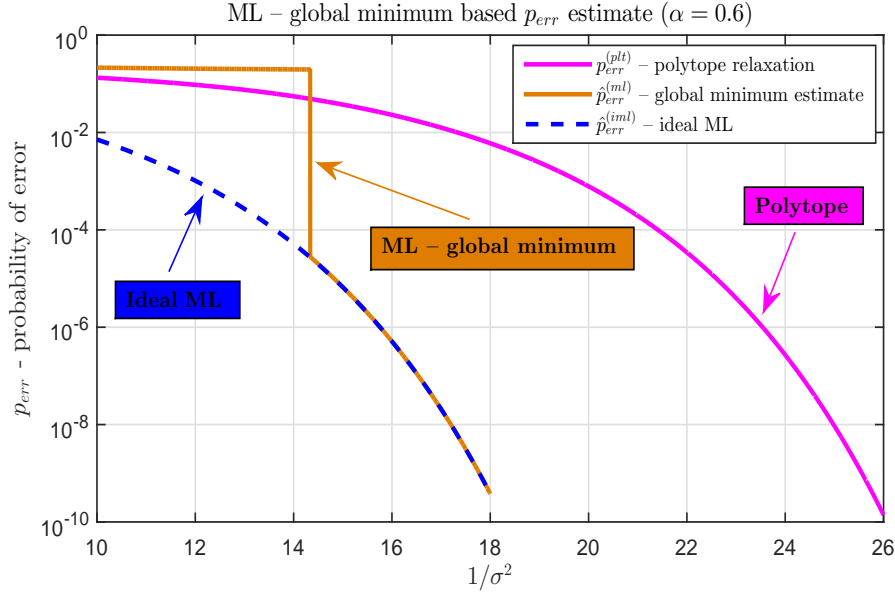,width=13.5cm,height=8cm}}
\caption{Comparison of $p_{err}$ as a function of $1/\sigma^2$; $\alpha=0.6$}
\label{fig:Rephasedprerrthml1}
\end{figure}

For concreteness we will set $\alpha=0.6$ (this is already very close to the $\alpha=0.5$ regimes where even in a noiseless scenario things are getting increasingly difficult for any known polynomial algorithm, for details in these directions see, e.g. \cite{StojnicISIT2010binary,Stojnicbinary16asym,DTbern,Wendel,DonTan09Univ}). We recall on the following theorem from \cite{Stojnicclupint19} that relates to the ML performance.
\begin{theorem}(ML -- RDT estimate \cite{Stojnicclupint19})
Assume the setup of Theorem 2 in \cite{Stojnicclupint19}. Then \begin{equation}
\lim_{n\rightarrow\infty}\mE\frac{\min_{\x\in{\cal X}}\|\y-A\x\|_2}{\sqrt{n}}=
\min_{c_1}\xi_{p}^{(ml)}(\alpha,\sigma;c_1)\geq \min_{c_1}\xi_{RD}^{(ml)}(\alpha,\sigma;c_1),\label{eq:thmdiscrd1}
\end{equation}
where
\begin{equation}
\xi_{p}^{(ml)}(\alpha,\sigma;c_1)\triangleq \lim_{n\rightarrow\infty}\frac{1}{\sqrt{n}}\mE \min_{\x\in {\cal X}}\max_{\|\lambda\|_2=1,\nu} \lambda^T\left ([A \v]\begin{bmatrix}\x_{sol}-\x\\\sigma\end{bmatrix} \right )+\nu((\x_{sol})^T\x-c_1), \label{eq:thmdiscrd2}
\end{equation}
and
\begin{equation}
  \xi_{RD}^{(ml)}(\alpha,\sigma;c_1)=\sqrt{\alpha}\sqrt{2-2c_1+\sigma^2}-\sqrt{2/\pi}\exp(-(\sqrt{2}\erfinv(-c_1))^2/2)).\label{eq:thmdiscrd3}
\end{equation}
\label{thm:mlthm1}
\end{theorem}
Moreover, one easily also has the following corollary.
\begin{corollary}(ML -- RDT estimate; stationary points) Let $\xi_{RD}^{(ml)}(\alpha,\sigma;c_1)$ be as in (\ref{eq:thmdiscrd3}). Then its stationary points are all solutions to
\begin{equation}
\frac{d \xi_{RD}^{(ml)}(\alpha,\sigma;c_1)}{dc_1} =-\frac{\sqrt{\alpha}}{\sqrt{2-2c_1+\sigma^2}}-\sqrt{2}\erfinv(-c).\label{eq:thmdiscrd4}
\end{equation}
Moreover, it also holds,
\begin{equation}
\frac{d^2 \xi_{RD}^{(ml)}(\alpha,\sigma;c_1)}{dc_1^2} =-\frac{\sqrt{\alpha}}{(2-2c_1+\sigma^2)^{3/2}}+\sqrt{\frac{\pi}{2}}\exp((\erfinv(-c))^2).\label{eq:thmdiscrd4}
\end{equation}
\label{cor:mlcor1}
\end{corollary}
In Figure \ref{fig:Rephasedprerrthml1} we show the ML estimates that one can obtain based on the above Theorem \ref{thm:mlthm1} and Corollary \ref{cor:mlcor1}. As mentioned earlier, throughout this section (and throughout the rest of the paper) we will assume that $\alpha=0.6$. In addition to the ML estimates we also show how they compare to the convex polytope relaxation ones and the so-called ideal ML (more on the definition and meaning of the ideal ML can be found in \cite{Stojniccluplargesc20}).
While the the above analysis produces the ML estimates that do approach the ideal ones they do so in a fairly high SNR regime, where the probability of errors are already of the order of $10^{-5}$. There is a very visible glitch that appears at around $1/\sigma^2=14.338$[db]. As emphasized through a thorough discussion in \cite{Stojnicclupint19}, this type of glitch indicates that things will not be as simple as one would like them to be. The effects of this type of observation in \cite{Stojnicclupint19} were rather mild. Here though things are substantially worse as the range of glitch affected SNRs is not only much wider but also falls exactly in the region where its presence is typically not very welcome. As the descriptions in Figure \ref{fig:Rephasedprerrthml1} hint, the source of the problems is the fact that the global optimum over $c_1$ in Theorem \ref{thm:mlthm1} and Corollary \ref{cor:mlcor1} does not reflect very well how things really behave.

\begin{figure}[htb]
\centering
\centerline{\epsfig{figure=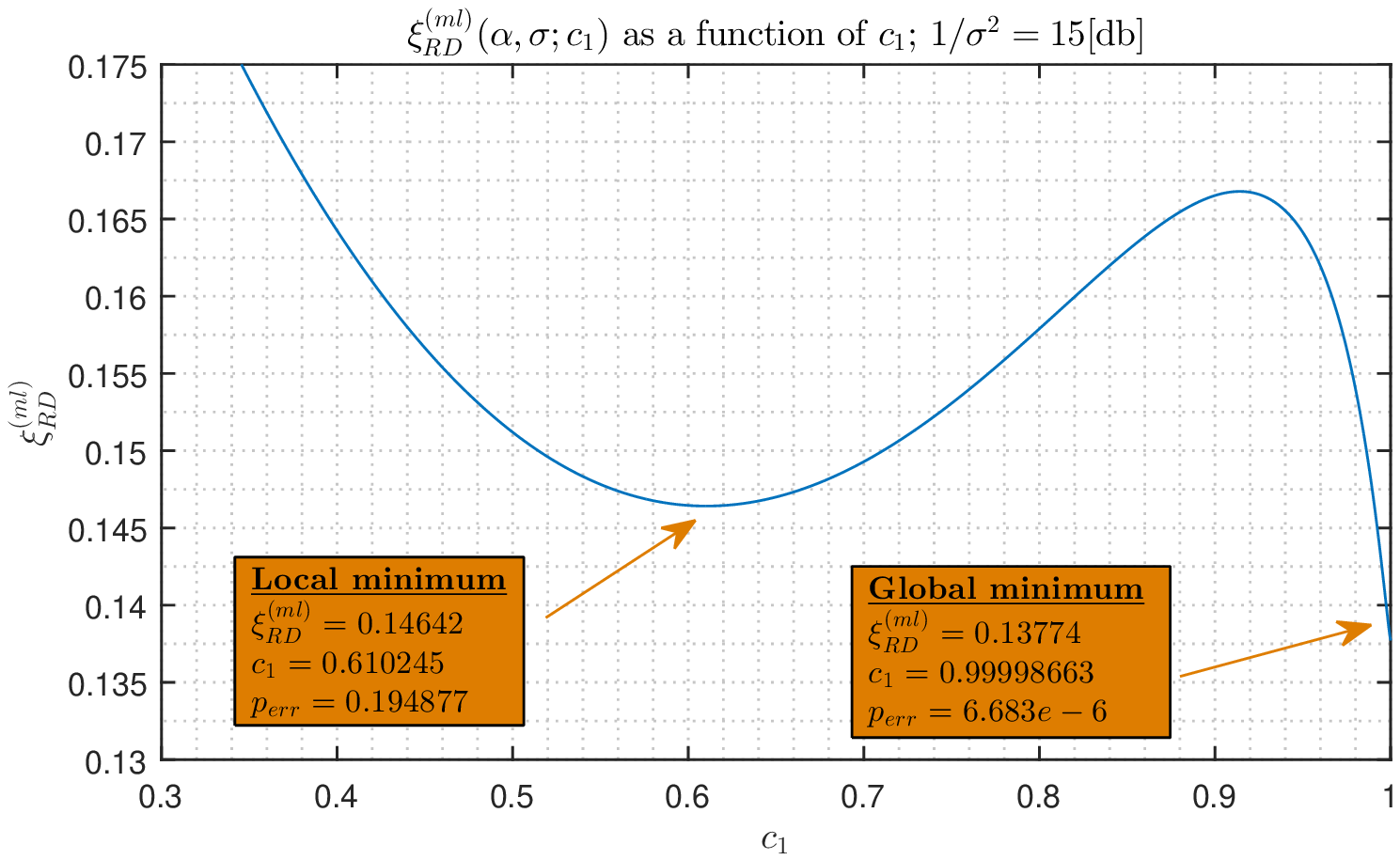,width=13.5cm,height=8cm}}
\caption{$\xi_{RD}^{(ml)}(\alpha,\sigma;c_1)$ as a function of $c_1$; $1/\sigma^2=15$[db]; $\alpha=0.6$}
\label{fig:Rephasedprerrthml2}
\end{figure}

\begin{figure}[htb]
\centering
\centerline{\epsfig{figure=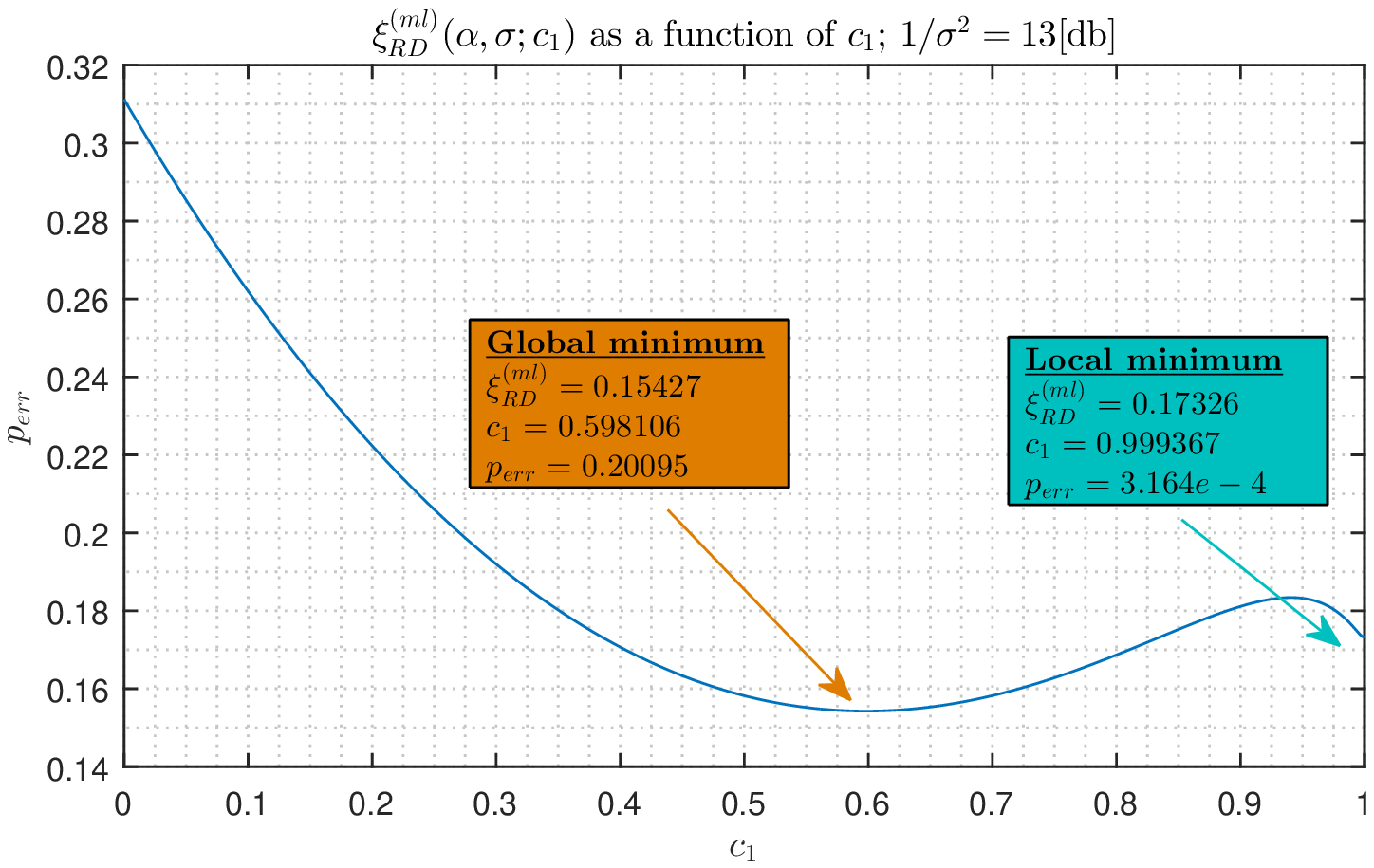,width=13.5cm,height=8cm}}
\caption{$\xi_{RD}^{(ml)}(\alpha,\sigma;c_1)$ as a function of $c_1$; $1/\sigma^2=13$[db]; $\alpha=0.6$}
\label{fig:Rephasedprerrthml3}
\end{figure}

\begin{figure}[htb]
\centering
\centerline{\epsfig{figure=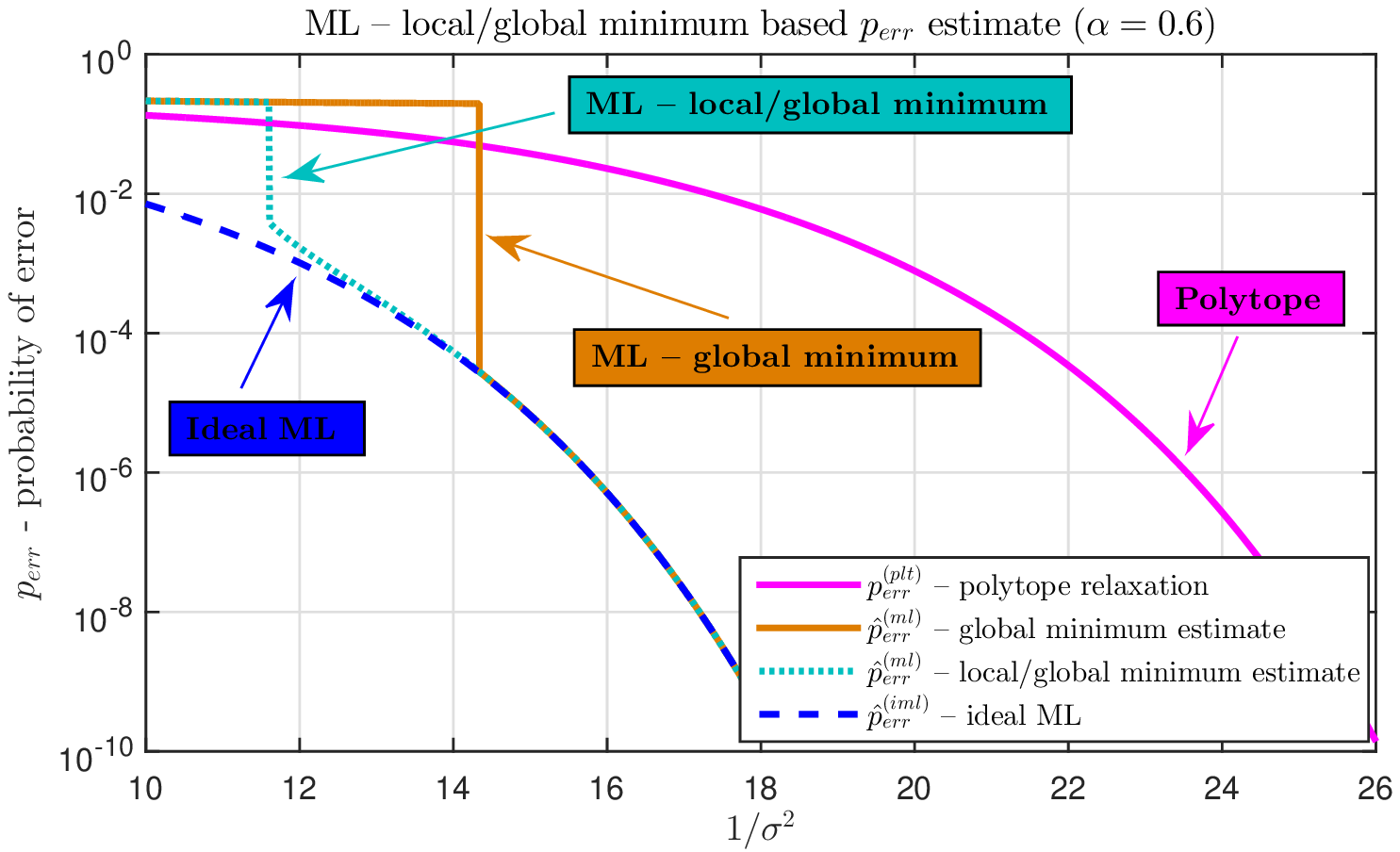,width=13.5cm,height=8cm}}
\caption{Comparison of $p_{err}$ as a function of $1/\sigma^2$; $\alpha=0.6$}
\label{fig:Rephasedprerrthml4}
\end{figure}

To give a bit of pictorial description we plot $\xi_{RD}^{(ml)}(\alpha,\sigma;c_1)$ as a function of $c_1$ in Figures \ref{fig:Rephasedprerrthml2} and \ref{fig:Rephasedprerrthml3}. In Figure \ref{fig:Rephasedprerrthml2} we fixed $1/\sigma^2=15$[db] and in Figure \ref{fig:Rephasedprerrthml3} we fixed $1/\sigma^2=13$[db]. One can now clearly see how the location of global and local minimum exchanged their places as the SNR goes down. Of course this has a rather profound effect on the value of the probability of error.
 However, if one uses a bit of so to say a bit of existing knowledge and assumes that the optimal $c_1$ should be rather large (i.e. close to $1$) and the probability of error rather small then a search for the local optimum provides a way better estimate. This is shown in Figure \ref{fig:Rephasedprerrthml4}. This is of course nice as one effectively neutralizes the glitch effect in a wide range of the most interesting SNR regimes. However, the main problem is likely to remain in place. Namely, the local optimum analysis does neutralize the glitch from the theoretical point of view but it doesn't provide a way as to how one can design practical algorithms that would have the same glitch neutralizing ability. Moreover, as the discussion in \cite{Stojnicclupint19} hinted, the appearance of local minima and their relevance in the ML analysis are quite connected to the structure and success of CLuP. That of course immediately raises a natural question: how would CLuP perform when $\alpha=0.6$ and given that the local minima appear to be of significant importance in the ML analysis in a wide range of SNRs. Before getting to the heart of the problem we below recall on the key fundamentals that stand behind the large scale CLuP design and discuss how a standard large scale CLuP implementation would fare in this low $\alpha$ regime.

\subsection{Analytical \bl{CLuP} foundation}
\label{sec:analfonclup}

To be able to discuss further what are the CLuP ultimate abilities when facing low $\alpha$ regimes we will need the following theorem from \cite{Stojniccluplargesc20} which we used as one of key foundations for establishing large scale CLuP.
\begin{theorem}(\bl{\textbf{Random dual}} -- stationary points \cite{Stojniccluplargesc20})
Consider the function $\xi_{RD,\gamma_1}(\alpha,\sigma;c_2,c_1,\gamma,\nu)$ given in the following way
\begin{equation}
\xi_{RD,\gamma_1}(\alpha,\sigma;c_2,c_1,\gamma,\nu)=-\sqrt{c_2}+\gamma_1(\sqrt{\alpha}\sqrt{1-2c_1+c_2+\sigma^2}+I_{22}-I_{1}+I_{21}-\nu c_1-\gamma c_2)-\gamma_1 r, \label{eq:LSclupthmeq1}
\end{equation}
with $I_{22}$, $I_{1}$, and $I_{21}$ defined as
\begin{eqnarray}
I_{22} &  = & \rho I_{22}(\gamma,\nu)+(1-\rho)I_{22}(\gamma,-\nu) \nonumber \\
I_{1}  &  = & \rho I_{1}(\gamma,\nu)+(1-\rho)I_{1}(\gamma,-\nu)  \nonumber \\
I_{21} &  = & \rho I_{21}(\gamma,\nu)+(1-\rho)I_{21}(\gamma,-\nu),
\label{eq:clupg16}
\end{eqnarray}
and $I_{22}(\gamma,\nu)$, $I_{1}(\gamma,\nu)$, and $I_{21}(\gamma,\nu)$ as in \cite{Stojniccluplargesc20}'s equation (12). Then its stationary points are all solutions to the following system of equations:
\begin{eqnarray}
  \frac{d\xi_{RD,\gamma_1}(\alpha,\sigma;c_2,c_1,\gamma,\nu)}{d\nu} & = & \rho\frac{dI(\gamma,\nu)}{d\nu}+(1-\rho)\frac{dI(\gamma,-\nu)}{d\nu}-c_1=0\nonumber \\
    \frac{d\xi_{RD,\gamma_1}(\alpha,\sigma;c_2,c_1,\gamma,\nu)}{d\gamma} & = &\rho\frac{dI(\gamma,\nu)}{d\gamma}+(1-\rho)\frac{dI(\gamma,-\nu)}{d\gamma}-c_2=0 \nonumber \\
      \frac{d\xi_{RD,\gamma_1}(\alpha,\sigma;c_2,c_1,\gamma,\nu)}{d\gamma_1} & = & \sqrt{\alpha}\sqrt{1-2c_1+c_2+\sigma^2}+I_{22}-I_{1}+I_{21}-\nu c_1-\gamma c_2-r=0 \nonumber \\
      \nu & = & -\frac{\sqrt{\alpha}}{\sqrt{1-2c_1+c_2+\sigma^2}} \nonumber \\
        \gamma_1 & = & \frac{1}{2\sqrt{c_2}(-\nu/2-\gamma)},
\end{eqnarray}
where $\frac{dI(\gamma,\nu)}{d\nu}$ and $\frac{dI(\gamma,-\nu)}{d\nu}$ are as given in \cite{Stojniccluplargesc20}'s equations (21)-(23), $\frac{dI(\gamma,\nu)}{d\gamma}$ and $\frac{dI(\gamma,-\nu)}{d\gamma}$ are as given in \cite{Stojniccluplargesc20}'s equations (25)-(27).
\label{thm:LSclupthm1}
\end{theorem}
As discussed in \cite{Stojniccluplargesc20}, if one wants to account for $c_2=1$ scenario the above needs a bit of tiny corrections. These corrections are rather trivial and as they don't play a major role in what follows we skip them. Also, as mentioned earlier, we will generally assume a certain degree of familiarity with the presentation from \cite{Stojniccluplargesc20}. We will also try to skip repetitions of many explanations already provided there. Along the same lines, we do recall that a little bit of familiarity with \cite{Stojniccluplargesc20} suggests that $\xi_{RD,\gamma_1}(\alpha,\sigma;c_2,c_1,\gamma,\nu)$ in the above theorem is the optimizing objective of the so-called \bl{\textbf{random dual}} and that
\begin{eqnarray}
c_2 & = & \|\x\|_2^2\nonumber \\
c_1 & = & (\x_{sol})^T\x. \label{eq:clupdefc2c1}
\end{eqnarray}
More on the importance and CLuP relevance of the random dual and all critical parameters including $c_2$ and $c_1$ can be seen through the analysis of the machineries from \cite{Stojnicclupint19,Stojnicclupcmpl19,Stojnicclupplt19,Stojniccluplargesc20}.

\subsection{Numerical results and practical \bl{CLuP} realization}
\label{sec:numpracclup}

\begin{figure}[htb]
\centering
\centerline{\epsfig{figure=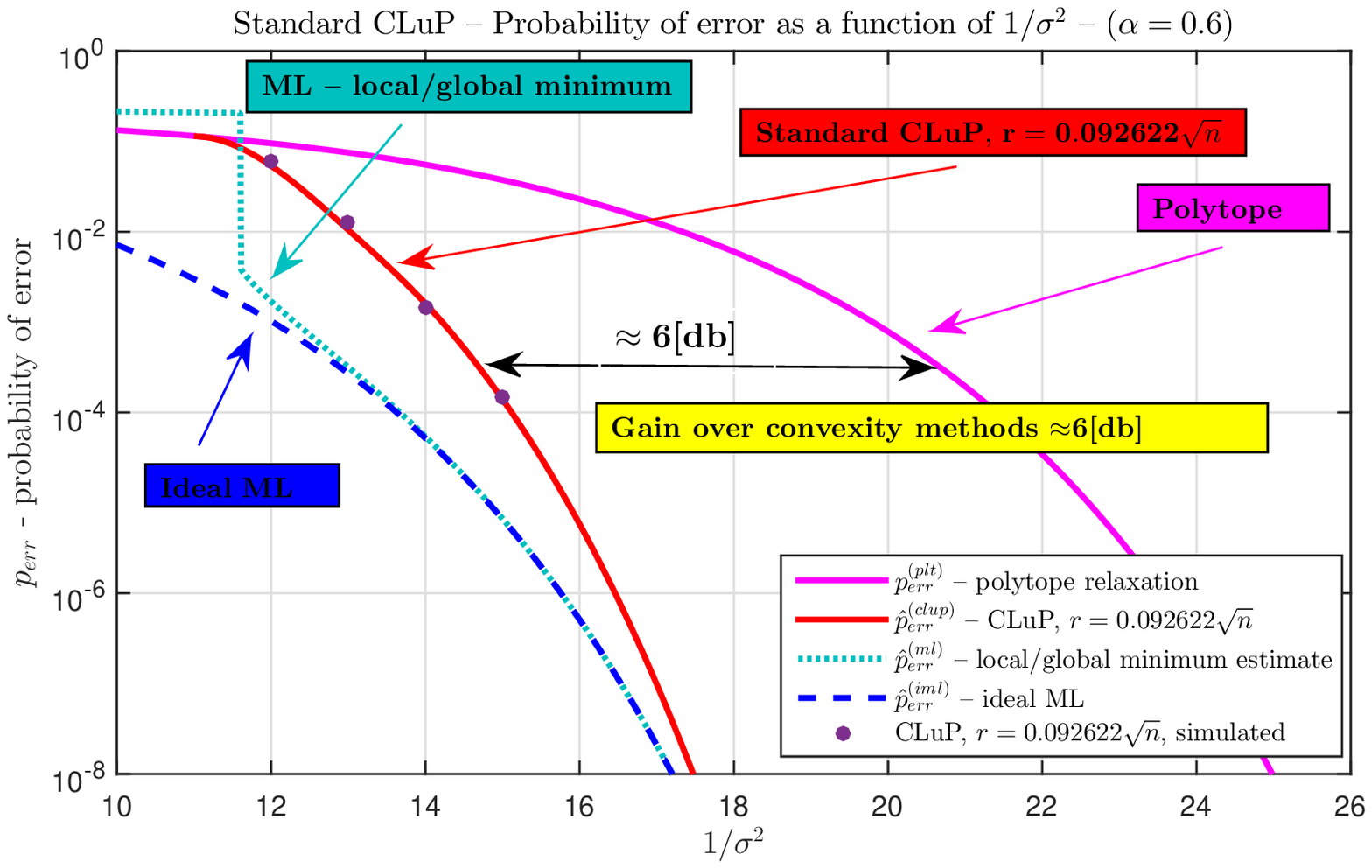,width=13.5cm,height=8cm}}
\caption{Standard CLuP, comparison of $p_{err}$ as a function of $1/\sigma^2$; $\alpha=0.6$, $r=0.092622\sqrt{n}$, $n=4000$}
\label{fig:Rephasedprerrthclup1}
\end{figure}

Based on the above theorem one can try to do both, provide a practical characterization of its performance and design an algorithm that would achieve such a performance. As hinted earlier things will not be as smooth as in \cite{Stojnicclupint19,Stojnicclupcmpl19,Stojnicclupplt19,Stojniccluplargesc20} though. So, we start with a little bit of warm up with a simple example. Namely, as mentioned earlier we choose $\alpha=0.6$ and set $r=0.092622\sqrt{n}$ (the choice is somewhat random but works reasonably well for what we want to highlight). In Figure \ref{fig:Rephasedprerrthclup1} we show the theoretical performance that one would expect based on Theorem \ref{thm:LSclupthm1} as well as the corresponding results that one gets through the numerical simulations.
We skip repeating all the details regarding the practical running of the large scale CLuP and instead mention that we ran everything conceptually exactly as explained in \cite{Stojniccluplargesc20}. Moreover, we complement the results presented in Figure \ref{fig:Rephasedprerrthclup1} with a solid set of numerical results presented in Table \ref{tab:tabrephasedclup1}. The numerical values in Table \ref{tab:tabrephasedclup1} are the exact values of all critical parameters obtained through both, the above mentioned theoretical considerations and the numerical experiments. The agreement between what the theory predicts and what the simulations give is rather solid.
\begin{table}[h]
\caption{\textbf{Theoretical}/\bl{\textbf{simulated}} values for $c_2$, $c_1$, $\hat{p}_{err}^{(clup)}$, and $r$ in Figure \ref{fig:Rephasedprerrthclup1} ($n=4000$)} \vspace{.1in}
\hspace{-0in}\centering
\footnotesize{
\begin{tabular}{||c||c||c|c||c|c||c|c||c|c||}\hline\hline
$ 1/\sigma^2 $[db] & $\hat{\gamma}_1\sqrt{n}  $ & $c_2$ & $c_2$ & $c_1$ & $c_1$ & $\hat{p}_{err}^{(clup)}$ & $\hat{p}_{err}^{(clup)}$ & $\frac{r}{\sqrt{n}}$ & $\frac{r}{\sqrt{n}}$ \\ \hline\hline
$12 $ & $\mathbf{2.5703 }$ & $\mathbf{0.7978 }$ & $\bl{\mathbf{0.8018 }}$ & $\mathbf{0.8148 }$ & $\bl{\mathbf{0.8076 }}$ & $\mathbf{5.3635e-02 }$ & $\bl{\mathbf{5.9175e-02 }}$ & $\mathbf{0.0926 }$ & $\bl{\mathbf{0.0935 }}$ \\ \hline
$13 $ & $\mathbf{2.0341 }$ & $\mathbf{0.8509 }$ & $\bl{\mathbf{0.8477 }}$ & $\mathbf{0.8943 }$ & $\bl{\mathbf{0.8900 }}$ & $\mathbf{1.0472e-02 }$ & $\bl{\mathbf{1.2599e-02 }}$ & $\mathbf{0.0926 }$ & $\bl{\mathbf{0.0926 }}$ \\ \hline
$14 $ & $\mathbf{1.5907 }$ & $\mathbf{0.8963 }$ & $\bl{\mathbf{0.8972 }}$ & $\mathbf{0.9335 }$ & $\bl{\mathbf{0.9345 }}$ & $\mathbf{1.6027e-03 }$ & $\bl{\mathbf{1.4640e-03 }}$ & $\mathbf{0.0926 }$ & $\bl{\mathbf{0.0930 }}$ \\ \hline
$15 $ & $\mathbf{1.2478 }$ & $\mathbf{0.9325 }$ & $\bl{\mathbf{0.9327 }}$ & $\mathbf{0.9593 }$ & $\bl{\mathbf{0.9595 }}$ & $\mathbf{1.3994e-04 }$ & $\bl{\mathbf{1.4462e-04 }}$ & $\mathbf{0.0926 }$ & $\bl{\mathbf{0.0926 }}$ \\ \hline\hline
\end{tabular}}
\label{tab:tabrephasedclup1}
\end{table}

Carefully looking at the results shown in Figure \ref{fig:Rephasedprerrthclup1} and Table \ref{tab:tabrephasedclup1} one can also observe that the standard CLuP, run with the above mentioned choice of parameters, achieves a fairly strong gain of approximately $6$[db] over the typical convex polytope relaxation. Moreover, such a gain is achieved with a fixed (SNR independent) choice for $r$. Still, being aware of our earlier success with CLuP (in particular its ability to approach the ideal ML performance) one naturally wonders can things be moved further in such a direction. The first question along such lines is in fact what would be the so-called ultimate CLuP performance. We refer to \cite{Stojnicclupint19,Stojnicclupcmpl19,Stojnicclupplt19,Stojniccluplargesc20} for the definitions and more complete discussions regarding the CLuP's ultimate performance. Here, we just briefly mention that the ultimate CLuP performance is achieved for a set of parameters that ensure that the resulting probability of error is minimal.
\begin{figure}[htb]
\centering
\centerline{\epsfig{figure=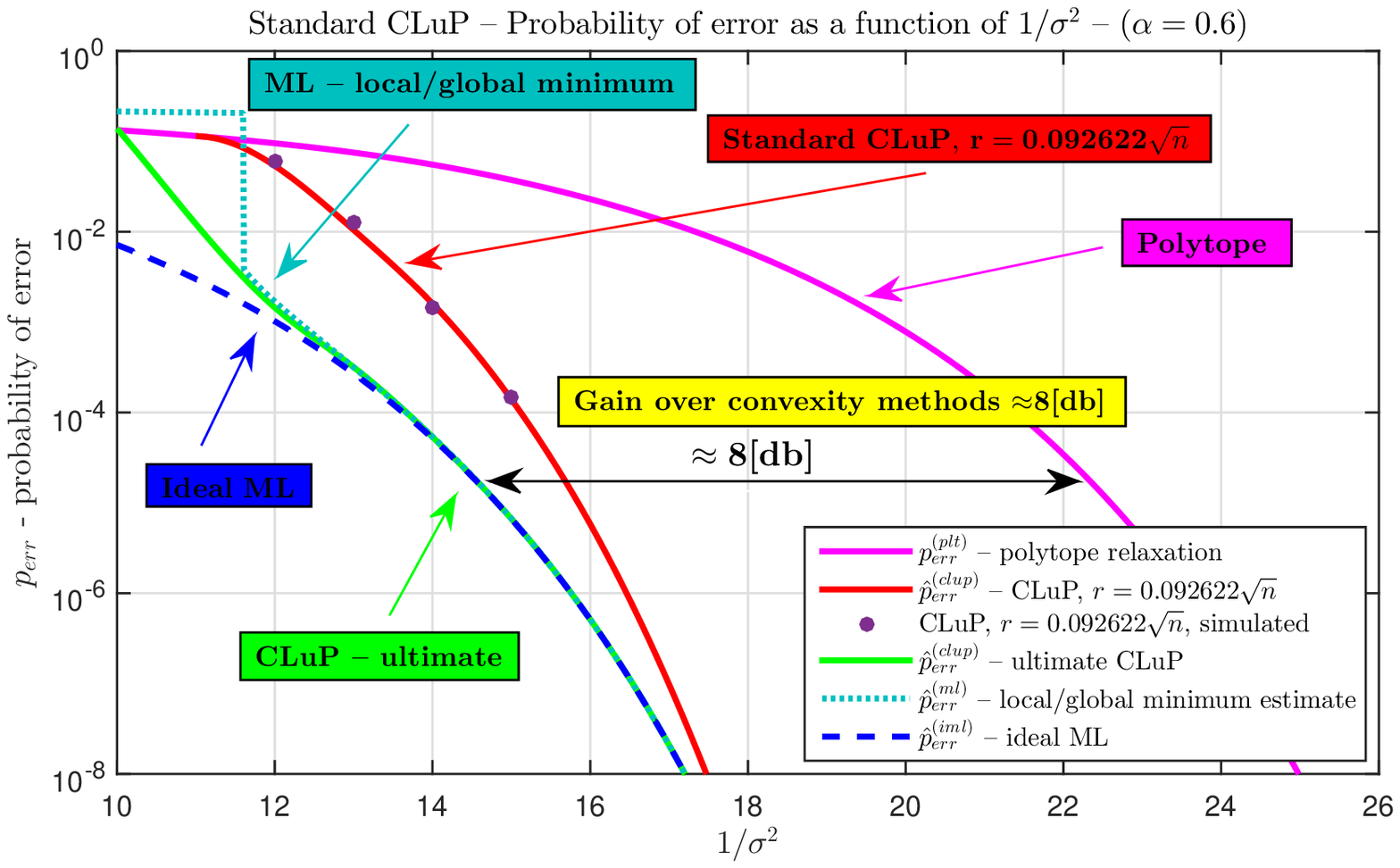,width=13.5cm,height=8cm}}
\caption{Ultimate CLuP, comparison of $p_{err}$ as a function of $1/\sigma^2$; $\alpha=0.6$}
\label{fig:Rephasedprerrthclup2}
\end{figure}
In Figure \ref{fig:Rephasedprerrthclup2} we show the theoretical predictions for the CLuP's ultimate performance that one can obtain through Theorem \ref{thm:LSclupthm1} and the machineries of \cite{Stojnicclupint19,Stojnicclupcmpl19,Stojnicclupplt19,Stojniccluplargesc20}. As can be seen from Figure \ref{fig:Rephasedprerrthclup2}, the ultimate CLuP achieves even stronger gain of approximately $8$[db] over the polytope relaxation. This is in a full agreement with what we hinted at in \cite{Stojnicclupint19}. Namely, in \cite{Stojnicclupint19}, we did mention on quite a few occasions that as $\alpha$ goes down running CLuP might not be as easy as in higher $\alpha$ regimes, but if successful the CLuP provides an even more substantial gain over the convex methods.
\begin{table}[h]
\caption{Numerical values for $r$ and $\hat{p}_{err}^{(clup)}$ that correspond to the data in Figure \ref{fig:Rephasedprerrthclup2} (green curve)} \vspace{.1in}
\hspace{-0in}\centering
\footnotesize{
\begin{tabular}{||c||c|c|c|c|c|c|c||}\hline\hline
$ 1/\sigma^2 $[db] & $10  $ & $11  $ & $12  $ & $13  $ & $14  $ & $15  $ & $16  $ \\ \hline\hline
$r/\sqrt{n}$ & $0.105  $ & $0.198  $ & $0.190  $ & $0.170  $ & $0.154  $ & $0.138  $ & $0.123  $ \\ \hline
$\hat{p}_{err}^{(clup)}$ & $1.38e-01  $ & $1.22e-02  $ & $1.45e-03  $ & $3.17e-04  $ & $5.36e-05  $ & $6.67e-06  $ & $5.12e-07  $  \\ \hline\hline
\end{tabular}}
\label{tab:tabclup2}
\end{table}
In Table \ref{tab:tabclup2} we show the explicit values for $r$ that achieve the ultimate CLuP performance. The corresponding probabilities of error are also given in parallel in Table \ref{tab:tabclup2} (they of course correspond exactly to the green curve in Figure \ref{fig:Rephasedprerrthclup2}).

\subsection{Obstacles on the path to ultimate CLuP}
\label{sec:obstultclup}

Of course, quite likely the most intriguing question regarding the above disparity between the standard and the ultimate CLuP performance is why it would be difficult to achieve the ultimate performance as $\alpha$ decreases and approaches regions around $0.5$. It is probably the easiest to explain the source of this difficulty through a concrete example. To that end let us consider the scenario where the SNR is $1/\sigma^2=14$[db] (as always, $\alpha=0.6$).
\begin{table}[h]
\caption{Values for $c_2$, $c_1$, $\nu$, $\gamma$, $\gamma_1$, $\hat{p}_{err}^{(clup)}$, and $r$ for stationary points at $1/\sigma^2=14$[db]; $r/\sqrt{n}=0.1544$} \vspace{.1in}
\hspace{-0in}\centering
\footnotesize{
\begin{tabular}{||c||c||c||c|c||c||c||c||}\hline\hline
Stat points & $c_2$ & $c_1$ & $\nu$ & $\gamma$ & $\gamma_1  $ & $\hat{p}_{err}^{(clup)}$ & $\frac{r}{\sqrt{n}}$  \\ \hline\hline
Stat point $1$ & $\mathbf{0.36708 }$ & $\mathbf{0.46120 }$ & $\mathbf{-1.11284 }$ & $\mathbf{1.14207 }$ & $\mathbf{-1.40913 }$ & $\mathbf{1.32889e-01 }$ & $\mathbf{0.1544}$ \\ \hline
Stat point $2$ & $\mathbf{0.97560 }$ & $\mathbf{0.89398 }$ & $\mathbf{-1.62420 }$ & $\mathbf{0.08536 }$ & $\mathbf{0.69656 }$ & $\mathbf{5.21666e-02 }$ & $\mathbf{0.1544}$ \\ \hline
\prp{\textbf{Stat point $3$}} &  $\prp{\mathbf{0.99992 }}$ & $\prp{\mathbf{0.99987 }}$ & $\prp{\mathbf{-3.87390 }}$ & $\prp{\mathbf{0.13117 }}$ & $\prp{\mathbf{0.27690 }}$ & $\prp{\mathbf{5.35544e-05 }}$ & $\prp{\mathbf{0.1544}}$ \\ \hline
\hline
\end{tabular}}
\label{tab:tabobstacles1}
\end{table}
We set the CluP ultimate performance achieving $r=0.1544\sqrt{n}$ and in Table \ref{tab:tabobstacles1} show the values for all critical parameters at the key stationary points of the random dual optimizing objective $\xi_{RD,\gamma_1}(\alpha,\sigma;c_2,c_1,\gamma,\nu)$. As can be seen there are three critical stationary points. The first one is trivially not much of a problem as the values of critical parameters are easy to circumvent (moreover if one insists on $\gamma_1$ being positive such a stationary point wouldn't even be permissible). The third stationary point is the desirable one. However, the appearance of the second one is exactly what might make problems and cause CLuP to struggle in achieving its ultimate performance and consequently approaching the ideal ML. It is not necessary the problem the mere appearance of this stationary point. For example, if it is in a way sufficiently far away from the desired one, one might be able to work around it and circumvent it. However, the key problem is actually exactly the fact that this unwelcome stationary point is sufficiently close to the desired one and it becomes a serious obstacle/trap for CLuP on its path towards achieving its ultimate performance and ultimately approaching the ML. The logic that we mentioned earlier when dealing with the ML performance, that one can rely on the local optimum and the fact that at the optimum $c_1$ is fairly large and in some regimes (for example when $1/\sigma^2=14$[db]) certainly above say $0.8$ or $0.9$ might not be enough to avoid being stuck in the second of these three stationary points. While this sounds a bit discouraging as it points out that the analytical structure of the underlying functions might be such that CLuP can't really do much, it in a way also provides a bit of hope that things might be so to say fixable. Namely, if one can find a solid starting $\x^{(0)}$ such that $c_1$ and $c_2$ are far away from the second stationary point and instead substantially closer to the third one, everything still may work. This is of course exactly the idea behind the \bl{\emph{\textbf{rephasing}}} concept that we briefly mentioned in \cite{Stojniccluplargesc20}. Overthere though that was more of a helpful tool rather than a necessity. On the other hand, overhere it is basically a must to get CLuP back on track and preserve its ML achieving ability.

\subsection{Achieving ultimate CLuP performance -- \bl{Rephasing}}
\label{sec:rephasing}

While the above mentioned logic of starting things with $\x^{(0)}$ that has favorable $c_1$ and $c_2$ properties in principle makes sense, it is not very clear that such $\x^{(0)}$'s are easy to find. In fact, it is conceivable that a priori finding such $\x^{(0)}$ might be as hard as finding the solution itself. Somewhat surprisingly CLuP itself does offer possibilities for finding favorable $\x^{(0)}$'s. Of course, to be able to observe that and exploit such opportunities one certainly needs to have a very strong understanding of the CLuP's fundamentals. As was the case above when we discussed the source of problems, we find it useful to discuss the remedies for the problems also through a particular example. Moreover, to be in a full agreement with the above example that helped introducing the problems, we will here also assume the same scenario in an attempt to resolve the problems. In other words, we will again assume $1/\sigma^2 =14$[db] and $\alpha=0.6$. Now, instead of looking at ultimate performance achieving $r$ we will here look again at a somehat random choice that we considered earlier, i.e. we will look at the scenario where $r=0.0926$.
\begin{table}[h]
\caption{Values for $c_2$, $c_1$, $\nu$, $\gamma$, $\gamma_1$, $\hat{p}_{err}^{(clup)}$, and $r$ for stationary points at $1/\sigma^2=14$[db]; $r/\sqrt{n}=0.0926$} \vspace{.1in}
\hspace{-0in}\centering
\footnotesize{
\begin{tabular}{||c||c||c||c|c||c||c||c||}\hline\hline
Stat points & $c_2$ & $c_1$ & $\nu$ & $\gamma$ & $\gamma_1  $ & $\hat{p}_{err}^{(clup)}$ & $\frac{r}{\sqrt{n}}$  \\ \hline\hline
 Stat point $1$ & $\mathbf{0.50526 }$ & $\mathbf{0.57484 }$ & $\mathbf{-1.23186 }$ & $\mathbf{0.93297 }$ & $\mathbf{-2.21874 }$ & $\mathbf{1.09000e-01 }$ & $\mathbf{0.0926}$ \\ \hline
\prp{\textbf{Stat point $2$}} & $\prp{\mathbf{0.89631 }}$ & $\prp{\mathbf{0.93352 }}$ & $\prp{\mathbf{-2.94733 }}$ & $\prp{\mathbf{1.14165 }}$ & $\prp{\mathbf{1.59070 }}$ & $\prp{\mathbf{1.60267e-03 }}$ & $\prp{\mathbf{0.0926}}$ \\ \hline
\hline
\end{tabular}}
\label{tab:tabrepahsing1}
\end{table}
In Table \ref{tab:tabrepahsing1} we show the values that all critical system parameters, $c_2$, $c_1$, $\nu$, $\gamma$, $\gamma_1$, $\hat{p}_{err}^{(clup)}$, and $r$, take at stationary points. One observes two stationary points with the first one being fairly far away from the second, desired one. Moreover, it is an easy exercise to repeat the so-called -- avoiding the lower stationary point -- arguments from \cite{Stojnicclupint19}'s section 3.2.6 and double check that once again one will not be stuck in the first stationary points with an overwhelming probability. Basically, the arguments of \cite{Stojnicclupint19}'s section 3.2.6 ensure that if one starts with a random initialization after the first iteration CLuP achieves a value of $c_2$ that is larger than the corresponding one of the lower stationary point and since $c_2$ progressively increases one is with overwhelming probability assured that the lower stationary point will indeed be circumvented (the same arguments assure that all the plots and simulated values in Figure \ref{fig:Rephasedprerrthclup2} are by no surprised exactly as shown in the figure).

\begin{figure}[htb]
\centering
\centerline{\epsfig{figure=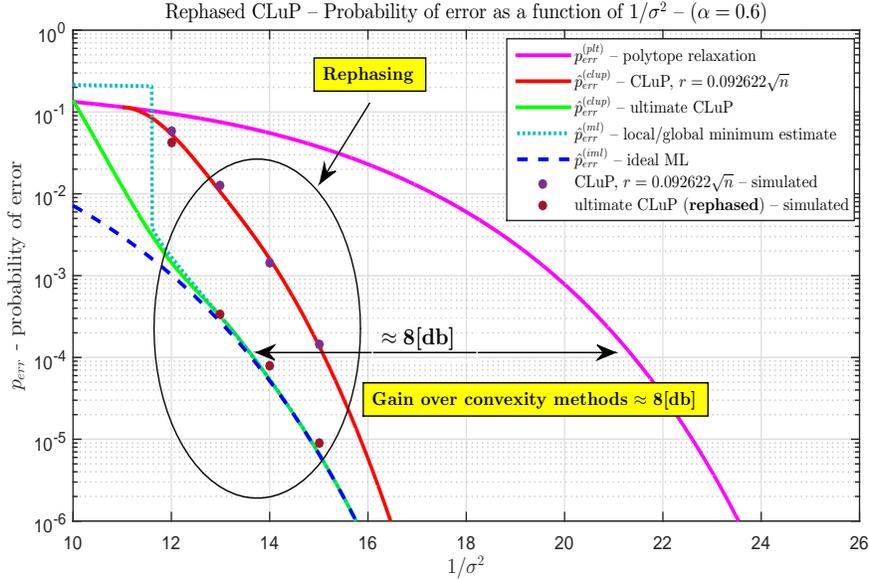,width=13.5cm,height=8cm}}
\caption{Ultimate CLuP (\textbf{rephased}), comparison of $p_{err}$ as a function of $1/\sigma^2$; $\alpha=0.6$}
\label{fig:Rephasedprerrthclup3}
\end{figure}

Now that one is guaranteed to reach the stationary point 2 from Table \ref{tab:tabrepahsing1}, one is also guaranteed to get to an $\x$ that has $c_2$, $c_1$, and their ratio far away from the the stationary point 2 from Table \ref{tab:tabobstacles1} and fairly close (so to say within the zone of attraction of the stationary point 3 from Table \ref{tab:tabobstacles1}). That means that we can rerun CLuP, or say its a large scale variant $\text{CLuP}^{r_0}$, with parameters that correspond to the stationary point 3 from Table \ref{tab:tabobstacles1} with $\x^{(0)}$ being the stationary point 2 from Table \ref{tab:tabrepahsing1}. We refer to such a mechanism as \bl{\textbf{CLuP rephasing}}. The resulting CLuP has two phases, phase 0 that corresponds to the discussion related to Table \ref{tab:tabrepahsing1} and phase 1 that corresponds to the discussion related to Table \ref{tab:tabobstacles1}. For the completeness, we also refer to the resulting two-phase CLuP as $\text{CLuP}^{r_1}$. Each phase is run as  $\text{CLuP}^{r_0}$ in a fashion explained \cite{Stojniccluplargesc20}, with $i_{max}=300$ and a two times larger $n$, i.e. this time we chose $n=4000$. Below we in Tables \ref{tab:tabclup4}, \ref{tab:tabclup5}, and \ref{tab:tabclup6}, show both, the theoretical and the simulated values for all critical system parameters for $1/\sigma^2\in\{13,14,15\}$[db] for complete $\text{CLuP}^{r_1}$, i.e. for both of its phases. The agreement between the theoretical predictions and the simulated values is again rather strong.
\begin{table}[h]
\caption{Rephasing -- \textbf{theoretical}/\bl{\textbf{simulated}} values for $c_2$, $c_1$, $\hat{p}_{err}^{(clup)}$, and $r$ ($n=4000$, $ 1/\sigma^2=13$[db])} \vspace{.1in}
\hspace{-0in}\centering
\footnotesize{
\begin{tabular}{||c||c||c|c||c|c||c|c||c|c||}\hline\hline
$ 1/\sigma^2 $[db] & $\hat{\gamma}_1\sqrt{n}  $ & $c_2$ & $c_2$ & $c_1$ & $c_1$ & $\hat{p}_{err}^{(clup)}$ & $\hat{p}_{err}^{(clup)}$ & $\frac{r}{\sqrt{n}}$ & $\frac{r}{\sqrt{n}}$\\ \hline\hline
$13 $ (phase 0) & $\mathbf{2.0341 }$ & $\mathbf{0.8509 }$ & $\bl{\mathbf{0.8477 }}$ & $\mathbf{0.8943 }$ & $\bl{\mathbf{0.8900 }}$ & $\mathbf{1.0472e-02 }$ & $\bl{\mathbf{1.2599e-02 }}$ & $\mathbf{0.0926 }$ & $\bl{\mathbf{0.0926 }}$ \\ \hline
$13 $ (phase 1) & $\mathbf{0.3869 }$ & $\mathbf{0.9976 }$ & $\bl{\mathbf{0.9975 }}$ & $\mathbf{0.9982 }$ & $\bl{\mathbf{0.9981 }}$ & $\mathbf{3.1720e-04 }$ & $\bl{\mathbf{3.4036e-04 }}$ & $\mathbf{0.1698 }$ & $\bl{\mathbf{0.1704 }}$ \\ \hline\hline
\end{tabular}}
\label{tab:tabclup4}
\end{table}
\begin{table}[h]
\caption{Rephasing -- \textbf{theoretical}/\bl{\textbf{simulated}} values for $c_2$, $c_1$, $\hat{p}_{err}^{(clup)}$, and $r$ ($n=4000$, $ 1/\sigma^2=14$[db])} \vspace{.1in}
\hspace{-0in}\centering
\footnotesize{
\begin{tabular}{||c||c||c|c||c|c||c|c||c|c||}\hline\hline
$ 1/\sigma^2 $[db] & $\hat{\gamma}_1\sqrt{n}  $ & $c_2$ & $c_2$ & $c_1$ & $c_1$ & $\hat{p}_{err}^{(clup)}$ & $\hat{p}_{err}^{(clup)}$ & $\frac{r}{\sqrt{n}}$ & $\frac{r}{\sqrt{n}}$\\ \hline\hline
$14 $ (phase 0)& $\mathbf{1.5907 }$ & $\mathbf{0.89631 }$ & $\bl{\mathbf{0.89718 }}$ & $\mathbf{0.93352 }$ & $\bl{\mathbf{0.93451 }}$ & $\mathbf{1.60e-03 }$ & $\bl{\mathbf{1.46e-03 }}$ & $\mathbf{0.0926 }$ & $\bl{\mathbf{0.0930 }}$ \\ \hline
$14 $ (phase 1)& $\mathbf{0.2769 }$ & $\mathbf{0.99992 }$ & $\bl{\mathbf{0.99991 }}$ & $\mathbf{0.99987 }$ & $\bl{\mathbf{0.99985 }}$ & $\mathbf{5.36e-05 }$ & $\bl{\mathbf{8.05e-05 }}$ & $\mathbf{0.1544 }$ & $\bl{\mathbf{0.1544 }}$ \\ \hline
\hline
\end{tabular}}
\label{tab:tabclup5}
\end{table}
\begin{table}[h]
\caption{Rephasing -- \textbf{theoretical}/\bl{\textbf{simulated}} values for $c_2$, $c_1$, $\hat{p}_{err}^{(clup)}$, and $r$ ($n=4000$, $ 1/\sigma^2=15$[db])} \vspace{.1in}
\hspace{-0in}\centering
\footnotesize{
\begin{tabular}{||c||c||c|c||c|c||c|c||c|c||}\hline\hline
$ 1/\sigma^2 $[db] & $\hat{\gamma}_1\sqrt{n}  $ & $c_2$ & $c_2$ & $c_1$ & $c_1$ & $\hat{p}_{err}^{(clup)}$ & $\hat{p}_{err}^{(clup)}$ & $\frac{r}{\sqrt{n}}$ & $\frac{r}{\sqrt{n}}$\\ \hline\hline
$15 $ (phase 0)& $\mathbf{1.2478 }$ & $\mathbf{0.93250 }$ & $\bl{\mathbf{0.93272 }}$ & $\mathbf{0.95933 }$ & $\bl{\mathbf{0.95952 }}$ & $\mathbf{1.40e-04 }$ & $\bl{\mathbf{1.45e-04 }}$ & $\mathbf{0.0926 }$ & $\bl{\mathbf{0.0926 }}$ \\ \hline
$15 $ (phase 1)& $\mathbf{0.3796 }$ & $\mathbf{0.99873 }$ & $\bl{\mathbf{0.99874 }}$ & $\mathbf{0.99923 }$ & $\bl{\mathbf{0.99924 }}$ & $\mathbf{7.19e-06 }$ & $\bl{\mathbf{8.82e-06 }}$ & $\mathbf{0.1358 }$ & $\bl{\mathbf{0.1357 }}$ \\ \hline\hline
\end{tabular}}
\label{tab:tabclup6}
\end{table}
We would also like to add that from Table \ref{tab:tabclup6} one can observe for $1/\sigma^2=15$[db] a slight difference between the ultimate performance achieving $r$ ($r\approx 0.138$ from Table \ref{tab:tabclup2}) and the value $r=0.1358$ that was used for simulations in this scenario. We chose a value a bit below the optimal one to ensure that even in finite dimensions one indeed approaches the ML. Since the value of $r$ is a bit below the one from Table \ref{tab:tabclup2} all other system parameters (including the probability of error) are a bit different from the ones obtained when the ultimate CLuP performance is achieved. In Table \ref{tab:tabclupx} we emphasize this and provide the exact values for $r$ that were simulated in each scenario as well as the corresponding probabilities of error.
\begin{table}[h]
\caption{Numerical values for $r$ and $\hat{p}_{err}^{(clup)}$ that correspond to the simulated data in Table \ref{tab:tabclup6} and Figure \ref{fig:Rephasedprerrthclup3}} \vspace{.1in}
\hspace{-0in}\centering
\footnotesize{
\begin{tabular}{||c||c|c|c||}\hline\hline
$ 1/\sigma^2 $[db]  & $13  $ & $14  $ & $15  $ \\ \hline\hline
$r/\sqrt{n}$ & $0.1698  $ & $0.1544  $ & $0.1358  $ \\ \hline
$\hat{p}_{err}^{(clup)}$  & $3.17e-04  $ & $5.36e-05  $ & $7.19e-06  $  \\ \hline\hline
\end{tabular}}
\label{tab:tabclupx}
\end{table}
Finally, in Figure \ref{fig:Rephasedprerrthclup3} we show the resulting plots that correspond to the above discussion and the values provided in Tables \ref{tab:tabclup4}, \ref{tab:tabclup5}, \ref{tab:tabclup6}, and \ref{tab:tabclupx}. In addition to the results that relate to $1/\sigma^2\in\{13,14,15\}$[db], we also in Figure \ref{fig:Rephasedprerrthclup3} show the results that one can obtain running $\text{CLuP}^{r_1}$ for $1/\sigma^2=12$[db] ($r/\sqrt{n}=0.0926$ in Phase 0 and $r/\sqrt{n}=0.1389$ in Phase 1). Since these results deviate from the theoretical predictions, we will in a separate section below discuss $1/\sigma^2=12$[db] scenario in a bit more details and uncover what are the sources of problems and how the problems can be remedied.

\subsection{Ultimate CLuP -- low SNR}
\label{sec:ultcluplowsnr}

Looking carefully at what we presented in the previous sections and in particular in Tables \ref{tab:tabclup4}, \ref{tab:tabclup5}, \ref{tab:tabclup6}, and \ref{tab:tabclupx} and Figure \ref{fig:Rephasedprerrthclup3} one can observe that the rephasing concept is very powerful and can bring the standard CLuP to actually approach yet again the exact ML in pretty much all relevant SNR regimes. Still, one wonders how would things work for SNRs that are below such regimes. Those regimes generally require a few bit more advanced considerations that we will discuss in separate papers. Here we would like to point out what happens at $1/\sigma^2=12$[db] (this value still falls in the range of the most interesting SNR regimes where the errors are not that large).
\begin{table}[h]
\caption{Values for $c_2$, $c_1$, $\nu$, $\gamma$, $\gamma_1$, $\hat{p}_{err}^{(clup)}$, and $r$ for stationary points at $1/\sigma^2=12$[db]; $r/\sqrt{n}=0.0926$} \vspace{.1in}
\hspace{-0in}\centering
\footnotesize{
\begin{tabular}{||c||c||c||c|c||c||c||c||}\hline\hline
Stat points & $c_2$ & $c_1$ & $\nu$ & $\gamma$ & $\gamma_1  $ & $\hat{p}_{err}^{(clup)}$ & $\frac{r}{\sqrt{n}}$  \\ \hline\hline
Stat point $1$ & $\mathbf{0.55891 }$ & $\mathbf{0.60232 }$ & $\mathbf{-1.19901 }$ & $\mathbf{0.81125 }$ & $\mathbf{-3.15855 }$ & $\mathbf{1.15262e-01 }$ & $\mathbf{0.0926}$ \\ \hline
\prp{Stat point $2$}  & $\prp{\mathbf{0.79781 }}$ & $\prp{\mathbf{0.81480 }}$ & $\prp{\mathbf{-1.61059 }}$ & $\prp{\mathbf{0.58751 }}$ & $\prp{\mathbf{2.57033 }}$ & $\prp{\mathbf{5.36347e-02 }}$ & $\prp{\mathbf{0.0926}}$ \\ \hline
\hline
\end{tabular}}
\label{tab:tabrephasinglowsnr12db1}
\end{table}

In Table \ref{tab:tabrephasinglowsnr12db1} we show the values of all critical systems parameters $c_2$, $c_1$, $\nu$, $\gamma$, $\gamma_1$, and $\hat{p}_{err}^{(clup)}$ for two key stationary points when $1/\sigma^2=12$[db] and $r/\sqrt{n}=0.0926$. This is precisely the scenario that corresponds to the standard CLuP with a particular radius choice that we discussed earlier (also as mentioned earlier, a discussion similar to the one from \cite{Stojnicclupint19}'s section 3.2.6 again ensures that CLuP avoids the lower stationary points (stationary point 1 in Table \ref{tab:tabrephasinglowsnr12db1}) and ends in the higher one (stationary point 2 in Table \ref{tab:tabrephasinglowsnr12db1})).

Table \ref{tab:tabrephasinglowsnr12db2} contains a set of results analogous to the ones presented in Table \ref{tab:tabrephasinglowsnr12db1}. The difference is that now $r/\sqrt{n}=0.1389$. One now observes the appearance of $4$ key stationary points. The values for all critical system parameters for each of these $4$ stationary points are given in the table. While stationary point 1 is again of no interest and far away from all others to pose any serious obstacle to CLuP's optimal convergence, the other three are fairly close to each other. In fact things are even a bit worse. Stationary points 2,3, and 4 are actually very close in norm-2 ($c_2$ value). However, stationary point 2's value for $c_1$ is far away from the corresponding ones of sstationary points 3 and 4. This may cause a serious problem when running CLuP. For example, if one were to run CLuP with $r/\sqrt{n}=0.1389$ it would be very hard to always avoid stationary point 2 and its fairly high probability of error. However, if one does the rephasing mechanism and first runs CLuP with parameters that correspond to the stationary point 2 from Table \ref{tab:tabrephasinglowsnr12db1} and then run CLuP with parameters that correspond to the stationary point 4 from Table \ref{tab:tabrephasinglowsnr12db2} it is highly likely that the stationary point 2 from Table \ref{tab:tabrephasinglowsnr12db2} will be circumvented. In fact as we will see a bit later not only will the stationary point 2 from Table \ref{tab:tabrephasinglowsnr12db2} be circumvented, the stationary point 3 from Table \ref{tab:tabrephasinglowsnr12db2} will be circumvented as well.

Similarly to Tables \ref{tab:tabrephasinglowsnr12db1} and \ref{tab:tabrephasinglowsnr12db2}, Table \ref{tab:tabrephasinglowsnr12db3} contains a set of results analogous to the ones presented in these two tables. The difference is that now $r/\sqrt{n}=0.1698$. This time we pay attention to $3$ key stationary points. Again, the first one will be trivially circumvented while the second one will be as well if one is able to get to the stationary point 4 from Table \ref{tab:tabrephasinglowsnr12db2} and utilize it as the starting point in running CLuP with $r/\sqrt{n}=0.1698$.

Finally, Table \ref{tab:tabrephasinglowsnr12db4} contains a set of results analogous to the ones presented in Tables \ref{tab:tabrephasinglowsnr12db1}, \ref{tab:tabrephasinglowsnr12db2}, and \ref{tab:tabrephasinglowsnr12db3} with $r/\sqrt{n}=0.1852$. There are again three key stationary points of interest and if one is able to get to the stationary point 3 from Table \ref{tab:tabrephasinglowsnr12db3} and utilize it as the starting point in running CLuP with $r/\sqrt{n}=0.1852$ then getting to the stationary point 3 from Table \ref{tab:tabrephasinglowsnr12db4} is within reach.
\begin{table}[h]
\caption{Values for $c_2$, $c_1$, $\nu$, $\gamma$, $\gamma_1$, $\hat{p}_{err}^{(clup)}$, and $r$ for stationary points at $1/\sigma^2=12$[db]; $r/\sqrt{n}=0.1389$} \vspace{.1in}
\hspace{-0in}\centering
\footnotesize{
\begin{tabular}{||c||c||c||c|c||c||c||c||}\hline\hline
Stat points & $c_2$ & $c_1$ & $\nu$ & $\gamma$ & $\gamma_1  $ & $\hat{p}_{err}^{(clup)}$ & $\frac{r}{\sqrt{n}}$  \\ \hline\hline
Stat point $1$ & $\mathbf{0.42111 }$ & $\mathbf{0.49600 }$ & $\mathbf{-1.10408 }$ & $\mathbf{1.01763 }$ & $\mathbf{-1.65488 }$ & $\mathbf{1.34779e-01 }$ & $\mathbf{0.1389}$ \\ \hline
Stat point $2$  & $\mathbf{0.94693 }$ & $\mathbf{0.62233 }$ & $\mathbf{-0.88540 }$ & $\mathbf{0.07386 }$ & $\mathbf{1.39308 }$ & $\mathbf{1.87970e-01 }$ & $\mathbf{0.1389}$ \\ \hline
Stat point $3$  & $\mathbf{0.92156 }$ & $\mathbf{0.87531 }$ & $\mathbf{-1.60115 }$ & $\mathbf{0.25560 }$ & $\mathbf{0.95573 }$ & $\mathbf{5.46724e-02 }$ & $\mathbf{0.1389}$ \\ \hline
\prp{Stat point $4$}  & $\prp{\mathbf{0.93010 }}$ & $\prp{\mathbf{0.95003 }}$ & $\prp{\mathbf{-2.53820 }}$ & $\prp{\mathbf{0.73723 }}$ & $\prp{\mathbf{0.97477 }}$ & $\prp{\mathbf{5.57122e-03 }}$ & $\prp{\mathbf{0.1389}}$ \\ \hline
\hline
\end{tabular}}
\label{tab:tabrephasinglowsnr12db2}
\end{table}

\begin{table}[h]
\caption{Values for $c_2$, $c_1$, $\nu$, $\gamma$, $\gamma_1$, $\hat{p}_{err}^{(clup)}$, and $r$ for stationary points at $1/\sigma^2=12$[db]; $r/\sqrt{n}=0.1698$} \vspace{.1in}
\hspace{-0in}\centering
\footnotesize{
\begin{tabular}{||c||c||c||c|c||c||c||c||}\hline\hline
Stat points & $c_2$ & $c_1$ & $\nu$ & $\gamma$ & $\gamma_1  $ & $\hat{p}_{err}^{(clup)}$ & $\frac{r}{\sqrt{n}}$  \\ \hline\hline
Stat point $1$ & $\mathbf{0.36239 }$ & $\mathbf{0.44968 }$ & $\mathbf{-1.06790 }$ & $\mathbf{1.12204 }$ & $\mathbf{-1.41233 }$ & $\mathbf{1.42782e-01 }$ & $\mathbf{0.1698}$ \\ \hline
Stat point $2$ & $\mathbf{0.96971 }$ & $\mathbf{0.92263 }$ & $\mathbf{-1.78866 }$ & $\mathbf{0.13862 }$ & $\mathbf{0.67189 }$ & $\mathbf{3.68350e-02 }$ & $\mathbf{0.1698}$ \\ \hline
\prp{Stat point $3$} & $\prp{\mathbf{0.97704 }}$ & $\prp{\mathbf{0.98332 }}$ & $\prp{\mathbf{-2.85736 }}$ & $\prp{\mathbf{0.59597 }}$ & $\prp{\mathbf{0.60746 }}$ & $\prp{\mathbf{2.13592e-03 }}$ & $\prp{\mathbf{0.1698 }}$ \\ \hline
\hline
\end{tabular}}
\label{tab:tabrephasinglowsnr12db3}
\end{table}

\begin{table}[h]
\caption{Values for $c_2$, $c_1$, $\nu$, $\gamma$, $\gamma_1$, $\hat{p}_{err}^{(clup)}$, and $r$ for stationary points at $1/\sigma^2=12$[db]; $r/\sqrt{n}=0.1852$} \vspace{.1in}
\hspace{-0in}\centering
\footnotesize{
\begin{tabular}{||c||c||c||c|c||c||c||c||}\hline\hline
Stat points & $c_2$ & $c_1$ & $\nu$ & $\gamma$ & $\gamma_1  $ & $\hat{p}_{err}^{(clup)}$ & $\frac{r}{\sqrt{n}}$  \\ \hline\hline
Stat point $1$ & $\mathbf{0.33734 }$ & $\mathbf{0.42958 }$ & $\mathbf{-1.05286 }$ & $\mathbf{1.17141 }$ & $\mathbf{-1.33472 }$ & $\mathbf{1.46203e-01 }$ & $\mathbf{0.1852}$ \\ \hline
Stat point $2$ & $\mathbf{0.98856 }$ & $\mathbf{0.94455 }$ & $\mathbf{-1.92116 }$ & $\mathbf{0.06772 }$ & $\mathbf{0.56323 }$ & $\mathbf{2.73560e-02 }$ & $\mathbf{0.1852}$ \\ \hline
\prp{Stat point $3$} & $\prp{\mathbf{0.99325 }}$ & $\prp{\mathbf{0.99408 }}$ & $\prp{\mathbf{-2.96639 }}$ & $\prp{\mathbf{0.36097 }}$ & $\prp{\mathbf{0.44706 }}$ & $\prp{\mathbf{1.50657e-03 }}$ & $\prp{\mathbf{0.1852}}$ \\ \hline
\hline
\end{tabular}}
\label{tab:tabrephasinglowsnr12db4}
\end{table}

\begin{figure}[htb]
\centering
\centerline{\epsfig{figure=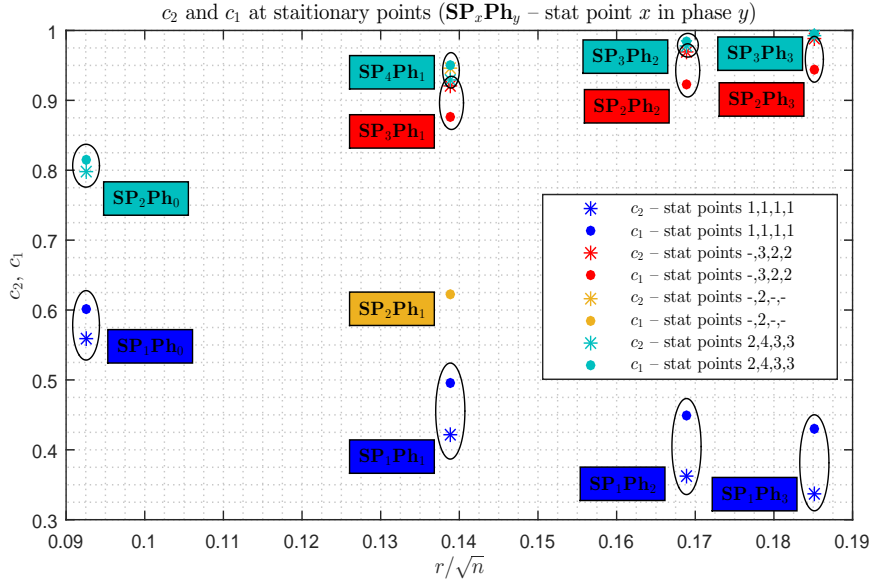,width=13.5cm,height=8cm}}
\caption{$c_2$ and $c_1$ at stationary points, $1/\sigma^2=12$[db]; $\alpha=0.6$; $r/\sqrt{n}\in\{0.0926,0.1389,0.1698,0.1852\}$}
\label{fig:ultcluplowsnr1}
\end{figure}

\begin{table}[h]
\caption{Rephasing -- \textbf{theoretical}/\bl{\textbf{simulated}} values for $c_2$, $c_1$, $\hat{p}_{err}^{(clup)}$, and $r$ ($n=8000$, $ 1/\sigma^2=12$[db])} \vspace{.1in}
\hspace{-0in}\centering
\footnotesize{
\begin{tabular}{||c||c||c|c||c|c||c|c||c||}\hline\hline
$ 1/\sigma^2 $[db] & $\hat{\gamma}_1\sqrt{n}  $ & $c_2$ & $c_2$ & $c_1$ & $c_1$ & $\hat{p}_{err}^{(clup)}$ & $\hat{p}_{err}^{(clup)}$ & $\frac{r}{\sqrt{n}}$ \\ \hline\hline
$12$ (Phase 0) & $\mathbf{2.57033 }$ & $\mathbf{0.79781 }$ & $\bl{\mathbf{0.79786 }}$ & $\mathbf{0.81480 }$ & $\bl{\mathbf{0.80592 }}$ & $\mathbf{5.363e-02 }$ & $\bl{\mathbf{6.150e-02 }}$ & $\mathbf{0.0926 }$ \\ \hline
$12$ (Phase 1)& $\mathbf{0.97477 }$ & $\mathbf{0.93010 }$ & $\bl{\mathbf{0.93341 }}$ & $\mathbf{0.95003 }$ & $\bl{\mathbf{0.95341 }}$ & $\mathbf{5.571e-03 }$ & $\bl{\mathbf{5.000e-03 }}$ & $\mathbf{0.1389 }$ \\ \hline
$12$ (Phase 2)& $\mathbf{0.60746 }$ & $\mathbf{0.97704 }$ & $\bl{\mathbf{0.97868 }}$ & $\mathbf{0.98332 }$ & $\bl{\mathbf{0.98502 }}$ & $\mathbf{2.136e-03 }$ & $\bl{\mathbf{1.875e-03 }}$ & $\mathbf{0.1698 }$ \\ \hline
$12$ (Phase 3)& $\mathbf{0.44706 }$ & $\mathbf{0.99325 }$ & $\bl{\mathbf{0.99391 }}$ & $\mathbf{0.99408 }$ & $\bl{\mathbf{0.99466 }}$ & $\mathbf{1.507e-03 }$ & $\bl{\mathbf{1.500e-03 }}$ & $\mathbf{0.1852 }$ \\ \hline
\hline
\ultclupcol{\textbf{$12$ (Ultimate)}} & $\ultclupcol{\mathbf{0.38907}} $ & \multicolumn{2}{ c|| }{$\ultclupcol{\mathbf{0.99711}}$} & \multicolumn{2}{ c|| }{$\ultclupcol{\mathbf{0.99630}}$} & \multicolumn{2}{ c|| }{$\ultclupcol{\mathbf{1.445e-03}}$}  & $\ultclupcol{\mathbf{0.1899 }}$ \\ \hline
\hline
\end{tabular}}
\label{tab:tabrephasinglowsnr12db5}
\end{table}
We refer to the above 4-phase  structure as $\text{CLuP}^{r_3}$. Obviously, there is really no restriction as to how many phases the rephasing process can have. Ideally, to decrease the computational complexity as much as possible one would tend to have as few phases as possible. However sometimes it may be beneficial to sacrifice a bit on overall computational complexity and add a phase or two to ensure a better fit of all concentration properties and an overall smoother and more reliable running. In general, we refer to a $k+1$-phase rephasing structure as $\text{CLuP}^{r_k}$. In Table \ref{tab:tabrephasinglowsnr12db5} we show the theoretical predictions as well as simulated results for a typical realization of the above discussed 4-phase  structure with $n=8000$. As can be seen the agreement between the theoretical predictions and the simulated values is very strong. Moreover, phase 4 is already very close to the theoretical ultimate CLuP performance. Still a few comments are in place. As the zones of attraction of almost all critical stationary points are not that far away from each other (see Figure \ref{fig:ultcluplowsnr1} where we show simultaneously the values of $c_2$ and $c_1$ at all critical stationary points) the passages in between them through which the CLuP should navigate are sometimes fairly narrow and one needs a very good concentration properties to ensure that the theoretical predictions practically indeed happen. Given that the differences (ratios) in probabilities of error if one gets to the right or wrong stationary point are sometimes of order of 100 times one has to have very strong concentrations of all relevant quantities. That basically means that about $70\%$ of success in achieving desired theoretically predicted concentrations (which happens for $n=4000$) might not be enough and is among the key reasons why the mean probability of error for $1/\sigma^2=12$[db] in Figure \ref{fig:Rephasedprerrthclup3} is closer to the standard CLuP than to the ultimate CLuP (the median though is very close to the ultimate CLuP prediction). However, as the dimensions increase this percentage increases as well and already dimensions $n$ of a few tens of thousands should typically be enough to achieve needed concentrations and the ultimate CLuP performance even for $1/\sigma^2=12$[db]. On the other hand, as mentioned earlier, as SNR drops further down to $11$[db] (or even below that) things are a bit more different. Since these SNRs are already outside the zone of low probability or errors and since they require a few additional considerations that go beyond the concepts that we presented here we will discuss them in a separate paper.

\section{Conclusion}
\label{sec:conc}

In our recent papers \cite{Stojnicclupint19,Stojnicclupcmpl19,Stojnicclupplt19} we presented some of the key concepts regarding the so-called CLuP algorithmic mechanism. As was clear already after the introductory considerations, the CLuP has quite a few features that are very favorable when one approaches solving some of the most challenging well-known optimization problems. To demonstrate CLuP power we in \cite{Stojnicclupint19,Stojnicclupcmpl19,Stojnicclupplt19} showcased its abilities when used for solving famous MIMO ML detection problem, one of the most fundamental algorithmic problems at the intersection of information theory, signal processing on the one side and statistics and machine learning on the other. Some of the main takeaways from \cite{Stojnicclupint19,Stojnicclupcmpl19,Stojnicclupplt19} were that CLuP approaches the exact ML performance while maintaining a very low computational complexity. Moreover, it turned out that CLuP's excellent predicated abilities happen in computationally the hardest regimes where the ratio of system dimensions are very unfavorable. In particular, already in \cite{Stojnicclupint19,Stojnicclupcmpl19,Stojnicclupplt19} it was clear that when it comes to the MIMO ML, CLuP can handle problem sizes of several hundreds in the hardest, so-called, $\alpha<1$ regime fairly well while basically maintaining its ability to achieve the exact ML.

In our companion paper \cite{Stojniccluplargesc20} we continued the discussion about CLuP and its computational complexity and showed how through a
\bl{\textbf{Random Duality Theory}} based analysis one can use it for attacking the so-called large scale problem instances where the dimensions are already of the order of a few thousands. In this paper, we present a continued discussion regarding the CLuP's behavior in low $\alpha$ regimes. While in \cite{Stojnicclupint19,Stojnicclupcmpl19,Stojnicclupplt19}  we chose $\alpha=0.8$ as a representative of the $\alpha<1$ regime, here we push things a bit further and discuss what happens as $\alpha$ approaches the $0.5$ zone where all known standard techniques start experiencing troubles even in the so-called noiseless (artificial) scenario within the MIMO ML detection.

We first observed that as $\alpha$ approaches $0.5$ things will not be as smooth as they were in \cite{Stojnicclupint19,Stojnicclupcmpl19,Stojnicclupplt19} for $\alpha=0.8$. To give a bit of a flavor as to what could be causing the troubles, we in parallel provided a discussion regarding the corresponding ML performance that one is ultimately trying to achieve. Through such a discussion we highlighted that the ML itself exhibits quite a few features that are different from say $\alpha=0.8$ regime. In particular, the underlying functions connected to the ML performance have an unfavorable behavior in a range of the SNRs that is much wider than the corresponding one for $\alpha=0.8$ in \cite{Stojnicclupint19,Stojnicclupcmpl19,Stojnicclupplt19}. Moreover, not only is this range much wider but it also contains quite a portion of the technically speaking the most relevant SNR range. Still, with a recognition that studying stationary points might be beneficial we were able to find a way to circumvent these troubles.

Given the inherent connection between the CLuP and the ML it was then expected that observations made regarding the ML might be beneficial in handling the CLuP as well. That was indeed true in two aspects: 1) it helped us recognize what the source of the CLuP's troubles can be as $\alpha\rightarrow 0.5$ and 2) it enabled us to design strategies that can help overcome such troubles. We first showed how one can utilize the basic CLuP to create algorithms that work universally well over pretty much the entire range of relevant SNRs while achieving performance substantially better than the typical convex relaxation methods. While such a performance is about $6$[db] better than the relaxation ones it is still on occasion about $2$[db] away from the ideal ML. To bridge the remaining gap we employed the so-called \bl{\emph{\textbf{rephasing}}} concept. We briefly mentioned this concept in \cite{Stojniccluplargesc20} as a state of the art tool which for regimes considered in  \cite{Stojniccluplargesc20} isn't necessarily needed but can be useful. Here though the concept is effectively the main tool that enabled us to bridge the $2$[db] gap between what the standard CLuP can achieve and what the desired ML performance is. The rephasing concept essentially assumes rerunning the basic CLuP in a particular way so that all the troubles caused by the unfavorable stationary points of the underlying so-called random dual optimizing objectives can be avoided. Of course, to do so, we heavily rely on the \bl{\textbf{Random Duality Theory}} itself, some of the key CLuP concepts from \cite{Stojnicclupint19,Stojnicclupcmpl19,Stojnicclupplt19}, and a long line of results that we created in our earlier work \cite{StojnicCSetam09,StojnicCSetamBlock09,StojnicISIT2010binary,StojnicDiscPercp13,StojnicUpper10,StojnicGenLasso10,StojnicGenSocp10,StojnicPrDepSocp10,StojnicRegRndDlt10,Stojnicbinary16fin,Stojnicbinary16asym}. A full utilization of all these machineries is also predicated on a favorable behavior of some of the underlying functions. That indeed happened to be the case and we were able to rephase CLuP so that it gets to the ML while pretty much maintaining the overall computational complexity. As was the case in \cite{Stojniccluplargesc20}, one here again works with the complexity per iteration of $mn$ operations (a single matrix/vector multiplication) which is for this type of problems theoretically minimal. The only difference compared to the standard large scale CLuP is that the overall number of iterations is now increased a couple of times due to the rephasing.

In addition to the presentation of the theoretical foundations that are behind the entire machinery, we also conducted quite a few numerical experiments and systematically presented a large set of results that we obtained through them. Quite a few of these results relate to the ML itself and many of course relate to the CLuP and essentially complement the corresponding theoretical discussions. We chose the large scale implementations that we discussed in \cite{Stojniccluplargesc20} and showed how they behave when facing the rephasing. We also observed a very strong agreement between the theoretical results and the simulated values. As expected for any of the considerations that we have done over the years within our random duality theory, the theoretical/simulated agreement is again so strong that it fairly often reaches the level of the fifth or even in some instances the sixth decimal for problem sizes of a few thousands. It is then expected that in large scale applications that reach dimensions of tens/hundreds of thousands or millions the agreement would be even better which is a particularly nice feature from the big data era prospective.

Of course, as expected there are many different ways how the rephasing can be done. We here presented a couple of basic ideas that are in a way tailored for the problem at hand, i.e. for the MIMO ML. However, the whole concept is in no way restricted to the MIMO ML detection. Quite contrary, it can actually be used for attacking any of the problems where the standard CLuP achieves success. Also, as discussed in \cite{Stojniccluplargesc20}, even when the rephasing is not needed it is often a helpful tool to ensure a better precision or if the dimensions are not that large to avoid a potential lack of concentration. As was discussed on quite a few occasions in \cite{Stojniccluplargesc20}, since in our introductory papers \cite{Stojnicclupint19,Stojnicclupcmpl19,Stojnicclupplt19} we chose the MIMO ML problem as the benchmark for showcasing CLuP's abilities, we here (as well as in \cite{Stojniccluplargesc20}) continued with the same practice. We found reusing the MIMO ML as the benchmark to be beneficial in doing both: 1) drawing the parallels and 2) emphasizing the differences between the basic CLuP and its more advanced structures presented here and in \cite{Stojnicclupplt19,Stojniccluplargesc20}. In separate papers we will discuss how all the CLuP concepts that we introduced in \cite{Stojnicclupint19,Stojnicclupcmpl19,Stojnicclupplt19} as well as in \cite{Stojniccluplargesc20} and here, can be utilized when facing different types of problems.

\begin{singlespace}
\bibliographystyle{plain}
\bibliography{cluprephasedRefs}
\end{singlespace}

\end{document}